\documentclass[11pt]{article}
\pdfoutput=1 
\usepackage{jheppub}
\usepackage{graphicx}

\usepackage[T1]{fontenc}
\usepackage{xcolor}
\usepackage{caption}
\usepackage{amsmath}
\usepackage{amsfonts}
\usepackage{mathrsfs}
\usepackage{comment}
\usepackage{subcaption,longtable,stmaryrd}
\usepackage{bigints}
\usepackage{multicol}
\usepackage{tikz,lipsum,lmodern}
\usepackage{dsfont}
\usepackage[most]{tcolorbox}
\usepackage{blindtext}
\usepackage{adjustbox}
\usepackage{tikz}
\usepackage{esvect}
\usepackage{float}
\usepackage{bbm}

\usepackage{parskip}

\DeclareFontFamily{U}{stix2bb}{}
\DeclareFontShape{U}{stix2bb}{m}{n} {<-> stix2-mathbb}{}

\NewDocumentCommand{\indicator}{}{\text{\usefont{U}{stix2bb}{m}{n}1}}
\usetikzlibrary{shapes, arrows.meta, positioning}

\tikzset{
  box/.style={
    rectangle,
    draw=black,  % Make the box visible
    thick,
    text width=12cm,
    minimum height=1.5cm,
    align=center,
    font=\large,
    rounded corners,
  },
  arrow/.style={thick, -{Latex[length=3mm]}},
}

\newcommand{\bea}{\begin{eqnarray}}
\newcommand{\eea}{\end{eqnarray}}

\newcommand{\I}{\mathscr{I}}

\title{Doubly-scaled planar ${\cal N} = 4$ SYM \& Carroll Holography}  

\author[1]{Arjun Bagchi,}
\author[2]{Prateksh Dhivakar,}
\author[3]{Alok Laddha,}
\author[4]{and Partha Paul.}
\author{\\}

\affiliation[1]{Indian Institute of Technology Kanpur, Kanpur 208016, India.\\}
\affiliation[2]{Department of Physics and Astronomy, University of Victoria, Victoria, BC V8W 2Y2, Canada.\\}
\affiliation[3]{Chennai Mathematical Institute
H1, SIPCOT IT Park, Siruseri, Kelambakkam 603103, India.\\}
\affiliation[4]{The Institute of Mathematical Sciences, C.I.T. Campus, Taramani, Chennai 600 113, India.\\ \&}
\affiliation[]{Homi Bhabha National Institute, Training School Complex, Anushakti Nagar, Mumbai 400094, India. \\ }
\emailAdd{abagchi@iitk.ac.in}
\emailAdd{pratekshd@uvic.ca}
\emailAdd{parthapaul@imsc.res.in}
\emailAdd{aladdha@cmi.ac.in}

\abstract{In their seminal paper, Okuda and Penedones (OP) put forward an intriguing proposal towards holography in asymptotically flat spacetimes (AFS). They showed that the flat space limit of 4 point bosonic tree-level string amplitude in AdS$_{5}\, \times\, S^{5}$ is dual to a specific double scaling limit of 4 point correlator in four dimensional (4d) $\mathcal{N} = 4$ super Yang-Mills (SYM) theory. In this paper, we show that the resulting boundary correlators can be used to define a 4d Carrollian conformal field theory (CFT) on future null infinity. This is done by mapping the set of doubly scaled correlators in SYM to null infinity, the conformal boundary of AFS. A Carroll CFT thus emerges in the infinite 't Hooft coupling limit of $\mathcal{N} = 4$ SYM. We extend the OP analysis to higher point functions and show how the Gram conditions on flat space amplitude put constraints on the double scaling limit of SYM correlators. We then use the soft factorization theorem for dilatonic string amplitude to write a recursion relation for the $({1}/{2})$-BPS sector of $\mathcal{N} = 4$ SYM in the double scaling limit. Finally, we show that the essential ideas underlying such a double scaling limit can be used as a tool-kit to build a class of Carrollian correlators, and hence define a Carrollian CFT, via certain integral transforms of flat space amplitudes of non-gravitational effective field theories.}

\begin{document}
\maketitle
\section{Introduction} 
The origins of flat space holography as a duality between quantum gravity in asymptotically flat spacetimes (AFS) and a non-gravitational quantum field theory~\cite{Polchinski:1999ry,Susskind:1998vk} arose right after Maldacena's discovery of the AdS/CFT correspondence~\cite{Maldacena:1997re,Witten:1998qj}. The seminal work in this direction was done by Polchinski~\cite{Polchinski:1999ry} and Giddings (and collaborators) ~\cite{Giddings:1999qu,Giddings:1999jq,Gary:2009ae}, who argued how the flat space S-matrix could be obtained via the large-radius limit of AdS correlators, which opened the window into the search for a gauge theory dual to flat space quantum gravity \cite{Bagchi:2023fbj,Bagchi:2023cen,Alday:2024yyj,Lipstein:2025jfj}.

\subsection*{Bottom-up v/s top-down}

However, right from its inception, the subject of flatspace holography has been dominated by advances in the bottom-up approach, where the non-gravitational S-matrix in Minkowski space is represented in terms of CFT correlators via the large-radius limit of the AdS S-matrix. This approach was pioneered by Penedones in \cite{Penedones:2010ue} and later developed systematically in \cite{ Fitzpatrick:2011jn,Fitzpatrick:2011hu,Fitzpatrick:2011dm,Raju:2012zr}. This has led to a beautiful crystallization of the approach to flat space S-matrix via correlators in Mellin space.

In other words, in the bottom-up approach we reconstruct the flat space amplitude in $d+1$ dimensional flat space in terms of conformal correlators on $d$ dimensional boundary of AdS$_{d+1}$. This has striking consequences for S-matrix bootstrap \cite{Paulos:2016fap,Paulos:2016but,Paulos:2017fhb}. A notable feature of this approach is that, the $d+1$ dimensional flat-space S-matrix is expressed in terms of (limit of a sequence of) conformal correlators defined on $\mathbb{R}^{1,d-1}$, which is the boundary of AdS$_{d+1}$ as opposed to Minkowski spacetime. This works as long as the correlators satisfy the Ward identities associated to Poincare group.

It may appear that such an approach then bears little resemblance to our conception of holography which is often articulated as a correspondence between  Quantum Gravity defined with respect to fixed conformal boundary and a dual non-gravitational theory which lives on this boundary as this would imply that the dual correlators live on the boundary of Minkowski space which is ${\I}^{+}\, \cup\, \hat{i}^{0}\, \cup\, {\I}^{-}$.\footnote{We denote the hyperbolic blow up of spatial infinity as $\hat{i}^{0}$ which supports the boundary representation of massive operators in the bulk spacetime, \cite{Laddha:2022nmj,H:2024cfo}. In fact $\hat{i}^{0}$ can also be used to construct boundary representation of massless fields as shown in the upcoming paper \cite{PV_Mukherjee:inprep}. This however is not a unique boundary representation of bulk operators as massive asymptotic operators have also been constructed in a blow up of time-like infinity \cite{Have:2024dff,Borthwick:2024skd,Liu:2025oom}. There have been attempts to understand massive bulk operators also on $\I^\pm$ \cite{Pasterski:2016qvg,Dutta:2026etj, Zheng:2026onv, Melton:2026tdw}.}
However, a moment of thought reveals that representation of a flat space S-matrix in terms of conformal correlators on boundary of AdS in fact does not invalidate the central premise of holography that a gravitational theory in $d+1$ dimensional space-time is dual to a non-gravitational QFT on a $d$ dimensional manifold.

Thus, there appear to be two rather widely differing approaches to the formulation of flat space holography, and in this paper, we show that the top-down approach advocated in \cite{Okuda:2010ym} can be ``integrated'' in the Carrollian approach to holography, which as we describe below has principally been formulated bottom up.

\subsection*{Celestial and Carrollian Holography}

Over the last decade, the bottom-up approach to flat space holography has been radically reformulated. The new insights have emerged thanks to the discovery of ``IR triangles'' which relate soft factorization theorems in gauge theories and gravity with asymptotic symmetries at the boundary of AFS \cite{He:2014laa,Lysov:2014csa,He:2014cra,Campiglia:2014yka,Avery:2015gxa,Campiglia:2016hvg,Strominger:2017zoo,Laddha:2017vfh,Banerjee:2020zlg,Guevara:2021abz,Strominger:2021mtt,Himwich:2021dau}. This approach is known as Celestial holography, where the central idea is to search for candidate {codimension two} CFTs on the celestial sphere at null infinity whose current algebra contains the algebra of asymptotic symmetries \cite{Strominger:2017zoo,Banerjee:2020zlg,Guevara:2021abz,Strominger:2021mtt,Himwich:2021dau,Banerjee:2021dlm,Raclariu:2021zjz,Pasterski:2021rjz,Pasterski:2021raf,Adamo:2021lrv,Costello:2022wso,Donnay:2023mrd,Melton:2024akx,Mol:2024etg,Ghorai:2026qaj,Ghorai:2025ebc,Zhu:2026ofh,Donnay:2022sdg,Pasterski:2021dqe}.

Celestial holography has recalibrated the bottom-up approach by bypassing the need to go to AdS and by systematically developing tools using which the flat space S-matrix for local QFTs can be represented as celestial correlators of a putative CFT \cite{Banerjee:2018gce,Pasterski:2017kqt}. In celestial CFT, the space of soft vacua is parameterized in terms of celestial Goldstone modes, associated to an infinite-dimensional tower of asymptotic symmetries, \cite{Donnay:2020guq, Puhm:2019zbl}.

However, despite many interesting advances, celestial holography remains, at its core, a bottom-up approach where 4d perturbative QFT observables (specifically the scattering amplitudes) are represented in terms of celestial CFT correlators. Notable exceptions are \cite{Banerjee:2020zlg,Banerjee:2021dlm, Adamo:2021lrv,Costello:2022wso,McLoughlin:2024ldp, Melton:2024akx,Mol:2024etg,Ghorai:2026qaj}, where asymptotic current algebra is used to classify the S-matrix in the MHV sector in gauge theories and gravity. In fact, several of these works go further and explicitly reconstruct the MHV amplitudes themselves as correlators of the 2d chiral current algebra \cite{Adamo:2021lrv,Costello:2022wso,Melton:2024akx,Mol:2024etg,Ghorai:2026qaj}. This approach does not require that a QFT which has a celestial dual be UV completed to string theory \cite{Mitra:2024ugt}, and hence it does not necessarily assume the existence of a quantum gravity theory for its existence. This is in stark contrast with AdS/CFT, which in its purest form exists as a duality between quantum gravity in AdS$_{5}\, \times\, S^{5}$ and ${\cal N}=4$ SYM theory \cite{Maldacena:1997re}. 

Carrollian holography has been the other popular approach to holography in asymptotically flat spacetime \cite{Bagchi:2025vri,Nguyen:2025zhg,Ruzziconi:2026bix}. After initially focusing on lower dimensions \cite{Bagchi:2010zz, Bagchi:2012cy,Bagchi:2012xr,Barnich:2012aw,Barnich:2012rz,Bagchi:2012yk, Bagchi:2014iea,Jiang:2017ecm}, there has been a recent resurgence of activities to connect the physically relevant 4$d$ AFS to 3$d$ Carrollian CFTs following \cite{Bagchi:2022emh,Donnay:2022aba,Donnay:2022wvx}. Carrollian theories are obtained in a vanishing speed of light limit of relativistic theories \cite{LevyLeblond,SenGupta:1966qer} and appear as symmetries on any null surface. Thus, in the context of AFS holography, these Carrollian CFTs naturally sit on the null boundary of AFS. There is an exact isomorphism between the bulk asymptotic symmetries given by the Bondi-van der Burg-Metzner-Sachs (BMS) group \cite{Bondi:1962px,Sachs:1962wk} and the Conformal Carroll group in a lower dimension \cite{Duval:2014uva}, which is analogous to the matching of bulk and boundary symmetries in AdS/CFT. This co-dimension one field theory seems more natural as a limit from AdS/CFT and it has been shown that this indeed is the case. In the bulk, the inverse of the radius of AdS acts as the speed of light on the boundary field theory and the flat space limit in the bulk turns into the vanishing speed of light limit \cite{Bagchi:2012cy}. This has recently been understood at the level of Witten diagrams as well \cite{Bagchi:2023fbj, Bagchi:2023cen}.

As in Celestial holography, the Carrollian approach to holography in AFS has also mostly been bottom up, where one has used symmetries to understand both sides of the putative duality, e.g. 4$d$ scattering has been understood in terms of correlation functions of 3$d$ Carroll CFT \cite{Bagchi:2022emh}, taking recourse to the modified Mellin transformation \cite{Banerjee:2018gce} and symmetries on both sides, in a manner similar to the Celestial story with usual Mellin transformations which link S-matrices to 2$d$ conformal correlators on the Celestial sphere \cite{Pasterski:2017kqt}. Another remarkable development has been the development of Carrollian partition functions to reproduce bulk physics \cite{Kim:2023qbl,Kraus:2024gso,Kraus:2025wgi,Isen:2026xoc}, which reproduces the above relation between scattering amplitudes and Carroll correlators.

But given that the construction of Carrollian theories is more systematic, e.g. through a $c\to0$ limit or an expansion about $c=0$ of a relativistic theory \footnote{One could also construct interacting Carroll theories using a lattice regulator \cite{Cotler:2024xhb}. However, in order to have a finite ``effective central charge'', one has to take an unconventional $N \to 0$ limit \cite{Cotler:2025dau}, which is the opposite of the limit discussed in this paper.}, Carrollian holography can perhaps be thought of as a tool to classify non-gravitational quantum field theories on the null boundaries of AFS, with the hope that the landscape of these theories would be essential in searching for a theory dual to string theory in flat space. The quest for finding an exact dual to AFS is still an elusive goal \footnote{See \cite{Bagchi:2024efs,Lipstein:2025jfj,Bagchi:2026emg} for initial efforts in constructing a flat hologram through the AdS$_4$/CFT$_3$ correspondence by using a Carrollian expansion of ABJM.}. 

In this paper, following \cite{Okuda:2010ym}, we take a bold step towards this. Our starting point is the original Maldacena proposal of AdS$_5$/CFT$_4$, following which a particular flatspace limit will land us up exactly on a Carrollian dual theory. On the field theory side, this is a double scaling limit which leads to an infinitely coupled sector of $\mathcal{N}=4$ Super Yang Mills. Thus, remarkably, the infinitely strongly coupled sector of SYM is dictated by Carrollian symmetries and can be thought of as a Carrollian CFT. The inverse 't Hooft coupling $\lambda^{-1} = g_{YM}^{-2} N^{-1}$ plays the role of an effective speed of light in this context. $\lambda \to \infty$ is equivalent to $c\to 0$ and this infinite coupling sector of SYM goes over to a strict $c=0$ Carrollian CFT. {\footnote{We emphasise that it is only the theory at infinite coupling that would be what corresponds to the exact $c=0$ Carrollian theory and we are not speaking about a Carrollian expansion about $c=0$ corresponding to a sequence of large $\lambda$ theories order by order. So the doubly scaled infinite coupling theory would necessarily be an Electric Carroll theory.}} We would like to stress that this is the first time that one is formulating a Carrollian CFT as an infinitely strong coupling sector 
of a relativistic CFT. We have reason to believe that the limit we advocate in this paper would work for theories beyond $\mathcal{N}=4$ SYM and in general for any large $N$ quantum field theory, thus providing a novel way of generating Carrollian theories. We elaborate below.

\subsection*{Focus of this paper: the Okuda-Penedones limit}
Since string theory in AdS$_{5}\, \times\, S^{5}$ is dual to the large $N$ limit of ${\cal N} = 4$ SYM theory on $\textrm{Mink}^{3,1}$, it is natural to expect that a systematic analysis of the dual of string scattering in AdS where the scattering takes place in an infinitesimal neighbourhood of a point, can be used to probe sectors of flat space holography which would then correspond to a holographic dual that lives on the conformal boundary of AdS space itself. 

In 2010, Okuda and Penedones in a remarkable paper \cite{Okuda:2010ym}, analyzed a specific double scaling limit of 4 point \emph{\textit{Lorentzian}} correlator in ${\cal N} = 4$ SYM theory on ${\mathbb{R}}^{1,3}$ and showed that it was dual to the tree-level flat space string amplitude with massless dilatons as external states at finite $s,t$. More in detail, they analyzed 4 point function $\langle\, O(x_{1})\, \dots\, O(x_{4})\, \rangle$ (or more precisely the connected correlators) of $\frac{1}{2}$ BPS scalar operators in terms of the two cross ratios
\begin{align}\label{cratios}
\sigma^2 := \frac{x_{13}^{2}x_{24}^{2}}{x_{12}^{2}x_{34}^{2}}, \quad
\sinh^2\rho =\, \frac{\textnormal{det}(x_{ij}^{2})}{4x_{13}^2 x_{24}^2 x_{12}^2 x_{34}^2},
\end{align} 
in the following limit
\begin{align}
N\, \rightarrow\, \infty, \quad g_{YM}^{2}\, =\, \frac{\lambda}{N}\, =\, \textrm{fixed}, \quad
\lim_{\lambda\, \rightarrow\, \infty}\, \lambda^{\frac{1}{4}}\, \rho\, =\, \textrm{fixed}.
\end{align}
This is the double scaling limit of SYM 4 point function along with a specific (large $N$) scaling of one of the two cross ratios. It is the infinite coupling limit with $g_{YM}$ fixed such that the four external points approach the light cone of a common reference point at a rate that scales with $N$, $\rho\,\sim\, \frac{1}{N^{\frac{1}{4}}}\, \rightarrow\, 0$.
\emph{Henceforth, we will refer to this double scaling limit as the OP limit.}  To the best of our knowledge, the OP limit has seldom been explored in its original form, but for a sampling of papers which review this limit we refer the reader to \cite{Alday:2014tsa,Goncalves:2014ffa,Alday:2018pdi,Li:2021snj} and its application to strong coupling limit of SYM correlators in the presence of defects, \cite{Alday:2024srr,Chen:2025cod}.

It was shown in \cite{Okuda:2010ym} that in this limit, the correlation functions admit a ``genus expansion'' in terms of $g_{YM}$ and that the leading order term is dual to the tree-level bosonic string amplitude in ${\mathbb{R}}^{1,9}$ with external states being massless dilatons.

The core idea underlying this proposal for flat space holography is striking, especially when placed in the context of many of the recent developments in the field. It says that a certain (perturbative) sector of flat space holography around 10$d$ Minkowski vacuum is a strongly coupled gauge theory defined on ${\mathbb{R}}^{1,3}$. On the other hand, the conformal boundary to 5$d$ Minkowski space-time is ${\I}^{+}\, \cup\, {\I}^{-}$ which is a Carrollian manifold and as a result there is an expectation that the mythical dual to flat space holography should be a Carrollian CFT. We take the first step in establishing a map between the OP limit of SYM theory and a family of correlation functions in a Carrollian CFT.

\subsection*{Outline of the paper}

This paper is organized as follows. In Sec.~\ref{review_4pt}, we review the embedding-space formalism and the seminal derivation presented by Okuda-Penedones, which uses AdS/CFT duality to map the flat space dilatonic string amplitude in ${\mathbb{R}}^{1,9}$ to the double scaling limit of a 4-point function of $\frac{1}{2}$-BPS operators in $\mathcal{N} =4$ SYM theory in ${\mathbb{R}}^{1,3}$. In Sec.~\ref{npt_gen}, we generalize this map to $n$-point amplitudes in ${\mathbb{R}}^{1,d + 5}$. We first consider the case where $d\, >\, n - 3$ ensures that the space of Mandelstam invariants is not subjected to the Gram constraints. In Sec.~\ref{gramimpose} we show how the proposal of Okuda and Penedones (for flat space holography) can be generalized to higher point amplitudes in ${\mathbb{R}}^{1,9}$. In particular, we derive the general integral transform that maps the flat space $7$-point amplitude $T_7$ to the boundary correlator ${\cal A}_{7}^{\textrm{flat}}$.

In Sec.~\ref{null_map}, we show that the doubly scaled SYM correlators can also be mapped (as distributions) on ${\I}^{+}$. Concretely, we map the OP kinematics of the correlator insertions onto the null infinity: working in global coordinates, we show that the OP scaling condition on the cross ratio $\rho$ forces the boundary insertions into two patches around $\tau=\pm \frac{\pi}{2}$ and we use this to rewrite the 4-point boundary correlator as a class of  correlators on $\I^+$ that satisfy global Ward identities of a Carrollian CFT.

In Sec.~\ref{soft_5pt}, we use the soft dilaton theorem for tree-level string amplitudes in flat space and derive a recursion relation that relates the 5 point function of the ${\cal N} = 4$ SYM theory in double-scaling limit to the 4-point function. We interpret this result as a factorization theorem for Carrollian correlators tied to the sub-leading soft dilaton theorem.

Sec.~\ref{opphicubed} illustrates the framework of the OP limit for the case of generic field theories beyond ${\cal N}\, =\, 4$ SYM with a simple example, viz. the massless $\phi^3$ theory in AdS$_4$, whose double-scaling limit we show defines a class of Carrollian correlators with unusual pole structure not previously seen in the Carrollian CFT literature. We show that on general grounds the Carroll operators that lead to such correlation functions would have unfamiliar fractional powers. We conclude in Sec.~\ref{conclusion} with a discussion of our results and some open questions. Appendix \ref{app_mandelstam} works out the map between Mandelstam angle variables and the conformal cross ratios invariant under the OP limit, used in Sec.~\ref{soft_5pt}.
\section{Revisiting the Okuda-Penedones limit}
\label{review_4pt}

\subsection{OP and ${\cal N}=4$ SYM 4 point correlators}
In this section, we first give a very brief overview of the  embedding space formalism which is used throughout the paper and then review the Okuda-Penedones limit for ${\cal N}=4$ SYM 4 point correlators. We only introduce the required notation and the reader can consult a number of review articles for more details. A few sampling of such articles can be found in \cite{Weinberg:2010fx,Costa:2011mg,Simmons-Duffin:2016gjk}. The basic idea behind this formalism is the realisation that $AdS_{d+1}$ can be embedded inside $ \mathbb{R}^{2,d} $ as a hyperboloid,
\begin{equation}\label{embed_ads}
X^2=-\left( X^0 \right)^2 - \left( X^{d+1} \right)^2 + \left( X^1 \right)^2 + \cdots + \left( X^{d} \right)^2=-L^2 \, .
\end{equation}
The parametrization of these embedding coordinates in terms of the global coordinates $ \left( \tau, \zeta, \vv{e} \right) $ is given by,
\begin{equation}\label{X_param}
X^0 = \frac{L}{\cos\zeta} \cos\tau, \, X^{d+1} =  \frac{L}{\cos\zeta}\sin\tau, \, X^a = \frac{L \sin\zeta}{\cos\zeta} e^a, \qquad a=1,\ldots,d,
\end{equation}
where $ \vv{e} $ is a $ d $-dimensional unit vector on $ S^{d-1}$. The $d$ dimensional boundary is a Lorentzian cylinder at $ \zeta = \frac{\pi}{2}$. Any point on the boundary can be associated with a null ray $[P]$ in the embedding space $\mathbb{R}^{2,d}$ which is an equivalence class of null vectors 
\begin{align}
P\, \sim\, \lambda\, P \quad \forall\, \lambda\, \in\, GL(1).
\end{align}
A convenient representative of the null ray is the embedding vector.
\begin{equation}\label{P_param}
P^0 = \cos\tau, \, P^{d+1} =  \sin\tau, \, P^a = e^a, \qquad a=1,\ldots,d.
\end{equation}
In terms of the usual Poincare co-ordinates, we have $ P_i = (P_i^+,P_i^-,P^\mu)=(1,x^2,x^\mu) $ and one obtains $ P_{ij} = -2 P_i \cdot P_j =  x_{ij}^2 $. 

We will now review the proposal of Okuda and Penedones towards a specific sector of flat space holography. By specific sector, we mean that it is determined by a fixed value of the axio-dilaton moduli on the boundary. As a result $g_{YM}$ remains a parameter of the boundary theory as we take the flat space limit. However, in a recent series of works Sen has argued that the holographic dual of string theory in flat space may have to treat the moduli differently. In that, observables (such as flat space S-matrix) associated to different points in the moduli space should belong to a single dual theory. see, e.g. \cite{Sen:2025bmj} for a beautiful discussion on this topic.

In Section (\ref{revofderivop4}) we review the derivation of the OP limit which relates the tree-level 4 point string amplitude with external dilaton states in $\mathbb{R}^{1,9}$ with the double scaling limit of a 4 point function in the boundary theory. But we begin by recalling some of the key aspects of the OP limit which, in our opinion, has not received enough attention in the literature so far.

\medskip

The fundamental ingredient of AdS/CFT is the identification of coupling constants in SYM theory and type IIB string theory in AdS$_{5}\, \times\, S^{5}$. 
\begin{equation}
\begin{gathered}
\left(\frac{L}{l_{s}}\right)^4\, =\, g^{2}_{YM}\, N,\, 
l_{s} = \sqrt{\alpha^{\prime}},\, g_{s} = \frac{\lambda}{4\pi N},\, 
\lambda\, =\, g_{YM}^{2}\, N,\, 
2 G^{10}_{N}\, =\, (2\pi)^{2}\, g_{s}^{2}\, (\alpha^{\prime})^{4}\nonumber\\
\implies\, l_{\textrm{pl}}^{8} = 2 \pi^{2} g_{s}^{2} l_{s}^{8},\,~ \textrm{and}\,~ N^{\frac{1}{4}} \sim\, \frac{L}{l_{pl}}
\end{gathered}
\end{equation}
where $(g_{s}, l_{s}^{2})$ are the string coupling and the string tension, respectively. $\lambda$ is the 't Hooft coupling and $G_{N}^{10}$ is the 10 dimensional Newton's constant. $L$ is the AdS radius.\footnote{The flat space limit is taken by taking $L\, \rightarrow\, \infty$ while keeping $l_{s}$ fixed and finite.}

 Okuda and Penedones considered the reduced 4 point function in 4 dimensional ${\cal N} = 4$ SYM theory of $\frac{1}{2}$ BPS Lagrangian density operator ${\cal O}$ which is dual to massless dilaton operator in AdS$_{5}\, \times\, S^{5}$. 
 We will denote the connected correlator as ${\cal A}_{4}$,
 \begin{align}
 {\cal A}_{4}(N, \lambda, \sigma, \rho)\, =\, \langle\, O(x_{1})\, \dots\, O(x_{4})\rangle_{c}.
 \end{align}
 And the reduced 4 point correlator as $\tilde{\cal A}_{4}$,\footnote{If we represent the boundary points as projective null cone inside $R^{2,4}$ then ${\cal A}_{4}$ is a section of the line bundle over the projective space. That is, if $L$ is a tautological line bundle over the projective null cone then ${\cal A}_{4}$ is a section of $L^{\otimes 4\triangle}$.  $\tilde{{\cal A}}_{4}$ on the other hand is a function on the projective space.}
\begin{align}
\tilde{{\cal A}}_{4}(N, \lambda, \sigma, \rho)\, :=\, \frac{\langle\, O(x_{1})\, \dots\, O(x_{4})\rangle_{c}}{\langle O(x_{1})\, O(x_{3})\rangle\, \langle O(x_{2})\, O(x_{4})\, \rangle}.
\end{align}
The boundary points $(x_{1}, x_{2})$ are in the past of $(x_{3}, x_{4})$ in ${\mathbb{R}}^{1,3}$, and $\rho,\, \sigma$ are defined in eqn.(\ref{cratios}). As discussed previously, $\rho$ measures the conformally invariant distance of the configuration of the four points from a light cone of a common reference point $x_{0}$.\footnote{The causal relation between the four points are specified by the $i\epsilon$ prescription in $x_{ij}^{2}$ which we suppress for brevity.} The OP limit is then defined as the infinite coupling limit of SYM theory such that, 
\begin{align}
\lim_{N\, \rightarrow\, \infty}\, N^{\frac{1}{4}}\, \rho\, =\, \textrm{finite}.
\end{align}
 Thus, the objects of interest are the limits of SYM 4 point connected and reduced correlators of $\frac{1}{2}$ BPS scalars,
\begin{align}\label{eq:a4_def}
\lim_{\textrm{OP}}\, \frac{1}{\lambda^{\frac{7}{4}}}\, {\cal A}_{4}(N, \lambda, \sigma, \rho)\, =:\, {\cal A}_{4}^{\textrm{flat}}(g_{YM}, \xi_{4}, \sigma)\nonumber\\
\lim_{\textrm{OP}}\, \frac{\tilde{{\cal A}}_{4}(N, \lambda, \sigma, \rho)}{ \lambda^{7/4}}\, =:\, {\cal F}_{4}(g_{YM}, \sigma, \xi_{4})
\end{align}
where 
\begin{align}\label{eq:xi4_main}
\xi_{4}\, :=\, \left( - \frac{\sigma}{1-\sigma}\sqrt{\lambda}\rho^2 \right)^{1/2}\,,
\end{align}
is the conformally invariant cross ratio which remains fixed in the double scaling limit.\\
The 4 point function, ${\cal F}_{4}(g_{YM}, \sigma, \xi_4)$ is a rather mythical object obtained as the strict $N = \infty$, $\lambda\, =\, \infty$ limit of SYM correlators. In spite of belonging to the infinite coupling sector of the theory, it admits a ``genus'' expansion in $g_{YM}$ for $g_{YM} << 1$.

\medskip

This can be seen as follows. The genus expansion of the connected correlator can be written as
\begin{align}
\langle O(x_{1})\, \dots\, O(x_{4})\, \rangle_{c}\, =\, \sum_{h=0}^{\infty}\, N^{-2h+2}\, {\cal A}_{4\, h}(\lambda, \sigma, \rho).
\end{align}
Since the two point function scales as $N^{2}$ in the large $N$ limit, we see that
\begin{align}
{\cal \tilde A}_{4}(N, g_{YM}, \sigma,\, \rho)\, =\, \sum_{h=0}^{\infty}\, N^{-2h-2}\, {\cal A}_{4\ h}(\lambda,\, \sigma,\, \rho)= \sum_{h=0}^{\infty}\, g_{YM}^{4h+4}\, \lambda^{-2h-2}\, {\cal A}_{4\ h}(\lambda,\, \sigma,\, \rho).
\end{align}
In the OP limit we then have, 
\begin{align}
{\cal F}_{4}(g_{YM}, \sigma, \xi_{4})&:=\, \sum_{h=0}^{\infty}\, g_{YM}^{4h+4}\, \lim_{\lambda\, \rightarrow\, \infty, \rho\lambda^{\frac{1}{4}}\, =\, \textrm{fixed}}\, \frac{\, \lambda^{-2h-2}\, {\cal A}_{4\ h}(\lambda, \sigma,\, \rho)\, }{\sigma^8 \lambda^{7/4}}\nonumber\\
\implies\, {\cal F}_{4}&=:\, \sum_{h=0}^{\infty}\, g_{YM}^{4h+4}\, {\cal F}_{4\, h}(\sigma,\, \xi_{4})=\, g_{YM}^{4}\, {\cal F}_{4\ 0}(\sigma, \xi_{4})\, +\, O(g_{YM}^{8}).
\end{align}
It was then shown in \cite{Okuda:2010ym} that the tree-level 10$d$ flat space string amplitude with massless dilatons as external states is related to ${\cal F}_{4\ 0}$ by an invertible transform
 \begin{align}\label{optrans}
i {\cal T}_{4}(s, t, l_{s}^{2})\, =\, \textrm{const.} \frac{\sqrt{stu}}{l_s^{11}s^7} \int_{-i\infty}^{+i\infty}\frac{d\xi_4}{2\pi i} \xi_4 \mathcal{F}_{4 \ 0}( -t/s, \xi_4) e^{\xi_4 l_s \sqrt{s}}
 \end{align}
for generic kinematical variables $s,t$. In the above equation ${\cal T}_{4}$ is the stripped amplitude in Type IIB string theory in flat spacetime. OP limit thus gives us an intriguing window into discovering a top-down approach to flat space holography at least in weak coupling regime of string theory.

 The OP limit can be contrasted with the conformal Regge limit which probes the high energy small angle elastic scattering of the bulk theory. To quantify their difference, it is useful to recast the OP limit as follows.

In the case of 4 point conformal correlators, the standard choice of the 2 cross ratios is the following.  
\begin{align}
u\, =\, \frac{x_{12}^{2} x_{34}^{2}}{x_{13}^{2} x_{24}^{2}}, \,\, v = \frac{x_{14}^{2} x_{23}^{2}}{x_{13}^{2} x_{24}^{2}}, \qquad u = z \overline{z}, \, v = (1 - z)(1 - \overline{z}),
\end{align}
where $z, \overline{z}$ are two independent \emph{real} cross ratios fixed in terms of Lorentzian distance on the boundary.  Okuda and Penedones choose $\sigma, \rho$ which are related to $z,\, \overline{z}$ via following relations. 
\begin{align}\label{eq:sigma_rho_def}
\sigma^{2}:=\, \frac{1}{z\overline{z}} \quad 
\sinh^{2}\rho:=\, \frac{1}{4} \sigma^{2}\, (z - \overline{z})^{2} 
\end{align}
The OP limit on the gauge theory side is now the following. 
We consider $N\, \rightarrow\, \infty$ with $g_{YM}$ fixed so that  $\lambda\, \rightarrow\, \infty$.   Now consider the kinematical regime so that 
\begin{equation}
    \sinh^2\rho \sim \frac{1}{N^{\frac{1}{2}}} \ \textrm{as}\ N\, \rightarrow\, \infty.
\end{equation}
Thus in the space of cross-ratios the OP kinematic regime is localized on a co-dimension 1 locus, 
\begin{align}
z - \overline{z} = \frac{1}{\sigma} \frac{a}{N^{\frac{1}{4}}}
\end{align}
where $a\, \sim\ O(1)$. The difference with conformal Regge limit is now immediate. In the latter case, once the Euclidean cross ratios are analytically continued around $z=1$ branch cut onto the second Riemann sheet, the limit is the one where
\begin{align}
z,\overline{z}\, \rightarrow\, 0 \quad \mbox{with} \quad \frac{z}{\overline{z}}\, =\, \textrm{fixed}. 
\end{align}
On the bulk side, the conformal Regge limit probes the high energy small angle string scattering and as a result it probes the flat space limit of AdS stringy S-matrix along with AdS curvature corrections. In the OP limit, the double scaling limit ensures that $\frac{l_{s}}{L},\, \frac{l_{pl}}{L}\, \rightarrow\, 0$ and $g_{YM} << 1$ implies that we can isolate tree-level finite $\alpha^{\prime}$ string amplitude via SYM correlators on the boundary.

We close this section with several remarks. 
\begin{itemize}
\item The OP limit of a SYM (reduced) correlator of $\frac{1}{2}$ BPS scalars generates correlation functions in the infinite coupling limit of the theory which admit a genus expansion with $g_{YM}$ being the expansion parameter. 
\item As anticipated in \cite{Okuda:2010ym}, to leading order in $g_{YM}$, one can relate the $n$ point tree-level dilatonic string amplitude in flat space to an $n$ point correlator in $N, \lambda\, \rightarrow\, \infty$ limit of ${\cal A}_{n}(N, g_{YM}, \{\sigma_{i}\})$ where $\{\sigma_{i}\}_{i=1}^{\frac{n(n-3)}{2}-1}$ are the set of independent Lorentzian cross ratios for $n$ insertions. 
\item Once the double scaling limit is taken, the resulting correlator on SYM side depends on $g_{YM}, \sigma, \xi = \xi(\rho, \sigma, N)$ and could be understood as a  correlator in $SU(\infty)$ gauge theory defined on 4 dimensional Minkowski space-time and more specifically on a particular co-dimension one singular locus in the space of cross-ratios which are defined with respect to 4 dimensional Minkowski metric.

In this paper we will prove that since the OP locus, 
\begin{equation}
    z - \overline{z} = \mathcal{O}\left(\frac{1}{N^{\frac{1}{4}}}\right)
\end{equation}
is a null manifold, we can use the resulting correlators to \emph{define} a set of Carrollian correlators,  
\begin{align}
{\cal A}^{\text{flat}}_{4}(g_{YM}, \xi, \sigma)\, =: 
\langle\,\tilde{O}(v_{1}, \hat{x}_{1})\, \dots\, \tilde{O}(u_{4}, \hat{x_{4}} )\rangle_{\textrm{connected}}
\end{align}
which then defines a Carrollian CFT. Similarly, ${\cal F}_{4}$ defines a reduced correlator in the Carrollian theory \footnote{We define the Carrollian correlators purely at $\mathscr{I}^+$ and thus, we have used the antipodal identification to identify $v = u$.},
\begin{align}
{\cal F}_{4}(g_{YM}, \xi, \sigma)\, =:\, \frac{\langle\,\tilde{O}(u_{1}, \hat{x}_{1})\, \dots\, \tilde{O}(u_{4}, \hat{x_{4}} )\rangle_{\textrm{connected}}}{ \langle\, \tilde{O}_{1} \tilde{O}_{3} \rangle\, \langle\, \tilde{O}_{2}\, \tilde{O}_{4}\, \rangle}
\end{align}
\end{itemize}
%%%%%%%%%%%%%%%%%%%%%%%%%%%%%%%%%%%%%%%
%%%%%%%%%%%%%%%%%%%%%%%%%%%%%%%%%%%%%%%

\subsection{OP limit from the bulk} \label{revofderivop4}
We now review the flat space limit of bulk amplitudes using the OP limit.

A popular approach for studying the flat space limit of bulk AdS amplitudes is via the Mellin representation of the amplitude. The amplitude is a function of Mellin variables $\delta_{ij}$ and the flat space limit is studied in the strong coupling limit such that 
\begin{align}
\frac{\delta_{ij}}{\sqrt{\lambda}} =: s_{ij} \quad \textrm{fixed}
\end{align}
where $\lambda = \left(\frac{L}{l_{s}} \right)^4 $ is the 't Hooft coupling. In this limit, the 4 point Mellin amplitude is related by an integral transform to the 4 point amplitude in the flat space $T_{4}(\{\beta\, s_{ij}\})$ as follows. 
\begin{align}
{\cal M}(\delta_{ij})
&\underset{\lvert\delta_{ij}\rvert\,\gg\,1}{\simeq}
\frac{R^{3-d}}{\Gamma\!\left(\frac{1}{2}\sum_{i=1}^{4}\Delta_i-\frac{d}{2}\right)}
\int_{0}^{\infty} d\beta\,
\beta^{\frac{1}{2}\sum_{i=1}^{4}\Delta_i-\frac{d}{2}-1}e^{-\beta}
T_{4}\!\left(-\frac{4\beta}{L^{2}}\,\delta_{ij}\right)
\nonumber\\
&\underset{d=4,\,\Delta_i=4}{=}
\frac{1}{L\,\Gamma(6)}\int_{0}^{\infty}d\beta\,
\beta^{5}e^{-\beta}
T_{4}\!\left(-\frac{4\beta}{L^{2}}\,\delta_{ij}\right) .
\end{align}
However, since our primary purpose is to relate the dual correlators in the flat space limit to Carrollian correlators, we will use the position space representation of the AdS amplitude as in Okuda and Penedones in \cite{Okuda:2010ym}.

In more detail, we review and generalize their derivation to $n$ point tree-level amplitudes. The key result in \cite{Okuda:2010ym} was the relationship of (a specific flat space limit) of the bulk amplitude with the SYM correlator via an integral transform that in the case of $n=4$ turns out to be Laplace transform. As we shall see in the following, the structural relationship between a class of flat space amplitudes in $\mathbb{R}^{1,9}$ with ${\cal F}_{n}$ remains the same $\forall\, n$ in the sense that the two are related by an invertible integral transform. However, the kernel of the transform from ${\cal T}_{n}\, \rightarrow\, {\cal F}_{n}$ varies with $n$.\footnote{In section \ref{soft_5pt}, we will see that this dependence on $n$ plays a key role in the analysis of soft factorization theorems for the bulk amplitude.} We then generalize this idea to the $n$ point amplitudes in section \ref{npt_gen}. Here we  simply summarize the main result.

Given the $n$ point string amplitude in $\mathbb{R}^{1,2d+1}$, $T_{n}$, the following integral transform generates a specific class of Lorentzian correlators in SYM theory in the strong coupling limit, 
\begin{align}\label{nptoptreeres}
{\cal F}_{n}(\xi_{n}, \{\sigma_{ij}\})\, =\, \int_{0}^{\infty} d\beta\, \beta^{\alpha(n,d)}\, {\cal K}_{n}(\beta, \xi_{n})\, T^{10}_{n}(l_{s}^{2}S = \beta, \frac{s_{ij}}{S}, g_{s}).
\end{align}
$T^{10}_{n}$ is the 10 dimensional flat space perturbative string amplitude with external states being massless dilaton scalars.
Here $S$ is a specific choice of Mandelstam invariant, which is taken as the energy scale and is transformed via ${\cal K}$. OP explicitly computed ${\cal K}_{n}$ for $n=4$ and showed that it is the kernel for the Laplace transform. $\xi_{n}$ is conjugate to $\beta$ and $\{\sigma_{ij}\}$ are the cross ratios that remain unchanged under the OP limit.

An interesting off-shoot of this proposal, in the context of flat space holography, is that it can lead to new insights in the ``bottom-up'' approach. Consider, for example, a massless scalar QFT in AdS$_{d+1}$. We can then use eqn.(\ref{nptoptreeres}) to derive the boundary representation of the bulk flat space amplitude $T(s,t)$ for the 4 point amplitude. The double scaling limit in this case is simply $L\, \rightarrow\, \infty, \rho\, L\, =\, \textrm{const}\, \sim\, \xi$. Hence in section \ref{opphicubed}, we analyze the ``OP limit'' for $\phi^{3}$ amplitude in $AdS_{4}$ with the coupling constant $c_{\phi}$.\footnote{The reason we put OP limit in quotes is since Okuda and Penedones were specifically interested in holography and hence their proposal was about flat space limit of perturbative string amplitude in terms of correlators of the boundary conformal field theory and this led to the infinite coupling limit of ${\cal N} = 4$ SYM correlators in the planar limit. In a garden variety QFT with no flavor index, the double scaling limit simply leads to a perturbative conformal correlator on the boundary.} More in detail, we analyze the tree-level 4 point $\phi^{3}$ amplitude and show that a double scaling limit in which 
\begin{align}
L\, \rightarrow\, \infty,\, \rho\, \sim\, L^{-1},\, c_{\phi} = \textrm{fixed}
\end{align}
leads to a specific boundary 4 point function that admits a perturbative expansion in $c$ and has a pole of order 7 in $\xi_4$. In section \ref{opphicubed}, we show that the resulting 4 point boundary correlators can be realized in terms of boundary insertions in a neighbourhood of $\pm\, \frac{\pi}{2}$ and can consequently be used to define a Carrollian correlator at null infinity.
%%%%%%%%%%%%%%%%%%%%%%%%%%%%%%%%%
%%%%%%%%%%%%%%%%%%%%%%%%%%%%%%%%%
\section{Flat space limit of $n$-point dilaton amplitude in string theory in $AdS_{D}\, \times\, S^{5}$}
\label{npt_gen}
In this section, we generalize the analysis of \cite{Okuda:2010ym} to derive a relationship between a generic $n$-point amplitude and a double scaling limit of the $n$-point boundary correlator in ${\cal N} = 4$ SYM theory. That is, we will show that given a flat space $n$-point amplitude in terms of one energy variable $\eta$ and the remaining angle variables $s^{0}_{ij}$, there exists an integral transform that relates this amplitude to a double scaling limit of the $n$-point correlator in ${\cal N} = 4$ SYM theory.

We start by considering the bulk amplitude in $\textrm{AdS}_{D = d+1}$ with $D >\, n-2$. In section \ref{gramimpose}, we will extend the formula to $D=n-2$ which then leads to the generalization of the OP limit of flat space amplitudes in $\mathbb{R}^{1,9}$ upto 7 points.

As before, we will always assume that the external states for the bulk amplitude are massless dilaton states having zero momenta along $S^{5}$. 
Let 
\begin{align}
{\cal A}_{n}\, =\, \int_{AdS_{D}^{\otimes\, n}}\, \prod_{j=1}^{n}\, d X_{j} \prod_{i=1}^n G_{B\partial}(P_{i}, X_{i})\, G^{D}_{\textrm{Ads}}(X_{1}, \dots, X_{n}).
\end{align}
where $X_{i}$ are the AdS co-ordinates in the embedding space ($X_{i}^{2} = - L^{2}$) and $P_{j}$ are the boundary insertions. $G_{\textrm{AdS}}$ is the amputated Green's function in $\textrm{AdS}_{D}$.\footnote{This Green's function is obtained by integrating over the unit $S^{5}$ since when the external states are operator insertions associated with dilatons which have zero mode along $S^{5}$, we get 
\begin{align}
G_{n}^{D}(X_{1}, \dots, X_{n})\, =\, \prod_{i=1}^{n}\, \int d\Omega_{i}\, Y_{\vec{0}}(\Omega_{i})\, G_{n}^{D+5}(\{ X_{i}, \Omega_{i}\,\})
\end{align}
In fact, in the flat space limit around a point $m$ of $AdS_{D}$, we have 
\begin{align}
T_{m}(\textrm{AdS}_{D}\, \times S^{5})\, =\, T_{m}(\textrm{AdS}_{D})\, \oplus\, R^{5}\, \sim\, R^{1,5+d}
\end{align}.}
We now consider the flat space limit of ${\cal A}_{n}$ as follows:
\begin{align}
X_{i}\, =\, X_{1} + Y_{i} \quad \forall\, i\, \in\, \{1, \dots, n\}
\end{align}
such that $Y_{1}\, =\, 0$ and $\vert Y_{i} \vert \ll L$. To leading order in $\frac{1}{L}$ the bulk amputated Green's function becomes the flat space amplitude as 
\begin{align}
{\cal A}_{n}&\approx\, \int_{\textrm{AdS}_{D}}\, d X_{1} \, G_{B\partial}(X_{1}, P_{1}) \, \int_{\mathbb{M}_{D}}\, \prod_{j=2}^{n}\, d Y_{j} \prod_{i=2}^n G_{B\partial}(P_{i}, X_1+Y_{i})\, G^{D}_{\textrm{flat}}(0,\, Y_{2},\, \dots, Y_{n})\nonumber\\
& \approx\, \left( \prod_{j=2}^{n} \frac{(-i)^{\Delta_j} C_{\Delta_j} L^{\Delta_j}}{\Gamma(\Delta_j)L^{ \alpha(d)}l_s^{\Delta_j}} \right)\, \int_{AdS_{D}}\, d X_{1}\, G_{B\partial}(X_{1}, P_{1})\, \int_{\mathbb{M}}\, \prod_{j=2}^{n}\, \bigg( d Y_{j} \int_{0}^{\infty}\, \frac{d\beta_{j}}{\beta_{j}}\, \beta_{j}^{\triangle_{j} }\, \nonumber \\
& \hspace{4cm} \times e^{-2 i\, \frac{\beta_{j} P_{j} \cdot (X_{1} + Y_{j})}{l_{s}}} \bigg) G_{\textrm{flat}}^{D}(0, Y_{2},\, \dots,\, Y_{n})
\end{align}
where
\begin{align}
\alpha(d) = \frac{d}{2} - \frac{1}{2} 
\end{align}

On defining, 
\begin{align}\label{fsmom}
k_{j}\, :=\, -\frac{2 \beta_{j}\, P_{j}}{l_{s}}
\end{align} 
we obtain the stripped flat space amplitude as the following integral transform of the bulk Green's function,\footnote{In the embedding space $P_{j}$ and hence $k_{j}$ is a vector in $R^{2,d}$, but as we will see below $k_{j}$ becomes a null momentum vector in the $d+1$ dimensional Minkowski space once we take the OP limit.}
\begin{align}\label{eq:scatttn_def}
\int_{\mathbb{M}}\, \prod_{j=2}^{n}\, d Y_{j} e^{-2 i\, \frac{\beta_{j} P_{j} \cdot Y_{j}}{l_{s}}}G_{\textrm{flat}}^{D}(0, Y_{2},\, \dots,\, Y_{n})\, = \, i \, T_{n}\left(k_{1}= - \sum_{i=2}^n k_i\right).
\end{align}
Momentum conservation is manifest since in the OP limit the bulk AdS correlator is localized in the tangent space around $X_{1}$ ensuring flat space translation invariance. As we will see in the following in eqn (\ref{fin_npt}), integral over $X_{1}$ will eventually lead to an integral over an overall energy scale with a fixed $n$ dependent integral kernel.

As a result, the leading order term in $\frac{1}{L}$ expansion can be written as, 
\begin{align}\label{Anflat}
{\cal A}_{n}&\approx\, \frac{1}{L^{n\, \alpha(d)}} \left(\frac{L}{l_s}\right)^{\Delta} \left( \prod_{j=1}^{n} \frac{(-i)^{\Delta_j} C_{\Delta_j} }{\Gamma(\Delta_j)}  \right) \, \prod_{i=1}^{n} \int_{0}^{\infty} d \beta_{i}\, \beta_{i}^{\triangle_{i}-1}\nonumber \\
&\hspace*{0.6in} \times \int_{AdS_{D}}\, d X_{1}\, e^{i X_{1} \cdot \left(\sum_{j=1}^{n} k_{j}\right)}\, i \, T_{n}\left(k_{1}= - \sum_{i=2}^n k_i\right),
\end{align} 
where $\Delta=\sum_{i=1}^n \Delta_i$. We can do the $X_{1}$ integral over the Poincare patch using the identity \cite{Okuda:2010ym}, 
\begin{align}
\int_{AdS_{D}} d X_{1} e^{ i X_{1} \cdot \left(\sum_{j=1}^{n} k_{j} \right)}\, =\,  (-i)^{1-\frac{d}{2}}\, \pi^{\frac{d}{2}}L^{d+1} \int_{0}^{\infty} \frac{d y}{y^{\frac{d+2}{2}}}\, e^{iy +  \frac{i}{2 y} \left(\frac{L}{l_{s}}\right)^{2} \sum_{i,j} \beta_{i} \beta_{j} P_{ij}}
\end{align}
where $P_{ij} = -2 P_{i} \cdot P_{j}$. Substituting the above expression in \eqref{Anflat}, we get
\begin{align}\label{Anflat1}
{\cal A}_{n}&\approx\, \frac{(-i)^{1-\frac{d}{2}} \pi^{\frac{d}{2}}}{L^{ n\, \alpha(d)\, -d-1}}\left(\frac{L}{l_s}\right)^{\Delta} \left( \prod_{j=1}^{n} \frac{(-i)^{\Delta_j} C_{\Delta_j} }{\Gamma(\Delta_j)}  \right) \,  \int_{0}^{\infty} \frac{d y}{y^{\frac{d+2}{2}}} \, e^{iy} \nonumber \\
& \times \, \prod_{i=1}^{n} \int_{0}^{\infty} d \beta_{i}\, \beta_{i}^{\triangle_{i}-1} e^{ \frac{i}{2 y} \left(\frac{L}{l_{s}}\right)^{2} \sum_{i,j} \beta_{i} \beta_{j} P_{ij}} \, i \, T_{n}\left(k_{1}= - \sum_{i=2}^n k_i\right)
\end{align} 
Eqn. (\ref{Anflat1}) is valid for fixed $D$ and $\forall\, n$.

We now assume that $d\, >\, n-3$. This implies that the $n$ vectors $P_{i}$ are linearly independent in $\mathbb{R}^{2,d}$ and the matrix $P_{ij}$ is non-degenerate.

Let then the spectrum of $P_{ij}$ be $\{\lambda_{0}, \dots, \lambda_{n-1}\}$ with the orthonormal basis of the eigen-vectors being $\vv{\psi}_{0}, \dots, \vv{\psi}_{n-1}$. The OP limit is defined by the following scaling of the eigenvalues,
\begin{align}
\lambda_{0}\, \sim\, \frac{1}{N^{\frac{1}{2}}}, \ \lambda_{i}\, \sim\, O(1) \quad \textrm{for}\ i\, \in\, \{1, \dots, n-1\} 
\end{align}
In order to analyze such a limit explicitly, we choose a basis in which 
\begin{align}
\vv{\psi}_{0} = \frac{1}{\sqrt{n}}(1, 1, \dots, 1)
\end{align}
Hence, 
\begin{align}\label{pijlambda0}
\sum_{j} P_{ij} = \lambda_{0}
\end{align}
We now parameterize $\vv{\beta}$ in terms of the orthonormal Eigen-basis of $P_{ij}$ as
\begin{align}
\vv{\beta} = \eta \, \vv{\psi}_{0} + \sum_{a=1}^{n-1}  \nu_{a}\, \vv{\psi}_{a} 
\end{align}
Hence,
\begin{align}
\sum_{i,j} \beta_{i} \beta_{j} P_{ij}\, =\,  \lambda_{0} \eta^{2} + \sum_{a=1}^{n-1} \lambda_{a} \nu_{a}^{2}\nonumber\\
 d^{n}\vv{\beta}\, =\, |J| \, d\eta\, \prod_{a=1}^{n-1} d \nu_{a}
\end{align}
where $J$ is the Jacobian determinant for the transformation $\vv{\beta} \to \vv{\psi}_{0}, \vv{\psi}_a$ with $|J|=1$. We can write the integral over $\beta_{i}$ as an integral over $\eta$ and $\nu_{a}$ as 
\begin{align}\label{etavnui}
\int \prod_{i=1}^n d \beta_{i} \beta_{i}^{\triangle_{i}-1}\, f(\vec{\beta}) =\, \int d\eta \, d^{n-1}\vv{\nu}\ \prod_{i}(\eta + \sum_{a=1}^{n-1}\nu_{a} \psi^i_a)^{\triangle_{i} - 1} f(\eta, \vec{\nu})\nonumber\\
=\, \int d\eta \, d^{n-1}\vv{\nu}\ \left(\, \eta^{\triangle - n} f(\eta, \vec{\nu})\, +\, \mathcal{O}(\nu_{a})\, \right)
\end{align} 
In the large $L$ limit 
\begin{align}\label{lambda0largeR}
\lambda_{0}\, \sim\, \frac{l_{s}^{2}}{L^{2}}
\end{align}
Eqns (\ref{pijlambda0}) and (\ref{lambda0largeR}), imply that 
\begin{align}
\lim_{L \, \rightarrow\, \infty} \sum_{j} P_{ij} = 0
\end{align} 
Thus, the choice of $\vv{\psi}_{0}$ implies momentum conservation in the flat space limit.

\medskip

Moreover, since $\lambda_{0}\, \rightarrow\, 0$ as $ L \, \rightarrow\, \infty$, the first term in eqn.(\ref{etavnui}) becomes a Gaussian integral over $\nu_{a}$ and can be evaluated exactly in the stationary phase approximation.\footnote{Since the saddle is at $\nu_{i} \approx\, 0$ we can in fact drop the second term inside eqn.(\ref{etavnui}).}
\begin{align}
\int d^{n-1}\vv{\nu}\, e^{\frac{i}{2y}\left(\frac{L}{l_{s}}\right)^2 \sum_{a} \lambda_{a} \nu_{a}^{2}}\, =\, \frac{1}{\sqrt{\textrm{det}^{\prime}(P)}}\, (\, 2\pi\, )^{\frac{n-1}{2}} e^{\frac{i (n - 1)\pi}{4}}\, \left(\frac{l_{s}}{L}\right)^{n-1}\, y^{\frac{n-1}{2}} 
\end{align}
with 
\begin{align}
\textrm{det}^{\prime}(P)\, =\, \prod_{a=1}^{n-1}\, \lambda_{a}\, .
\end{align}
Finally, only the $y$ integral remains which can be obtained in terms of the Bessel function as
\begin{align}
\int_{0}^{\infty} d y \, y^{\frac{n-d-3}{2}}\, e^{i y - \frac{i}{y}\, \xi_{n}^{2}\eta^{2}}\, =\, 2 e^{\frac{i(n-d-1)\pi}{4}}(\xi_{n}\eta)^{\frac{n-d-1}{2}}\, K_{\frac{n-d-1}{2}}(2\xi_{n}\eta)
\end{align}
where we have defined
\begin{equation}\label{eq:xin2_def}
    \xi_{n}^{2} = - \lim_{(L/l_s)^2 \to \infty}\frac{1}{2} \left( \frac{L}{l_s}\right)^2 \lambda_{0}\, .
\end{equation} 
Putting all the pieces together, in the large $L$ expansion (and as $\lambda_{0}\, \rightarrow\, \frac{1}{L^{2}}$), we can write the leading order term (in the large $L$ expansion with OP scaling of one of the conformal cross ratios) of the boundary correlator ${\cal A}_{n}$ in terms of flat space S-matrix $T_{n}$, 
\begin{align}\label{fin_npt0}
{\cal A}_{n}(\xi_{n}, \vec{\sigma})\, \approx\, \frac{\mathcal{N}_n}{\sqrt{\textrm{det}^{\prime}(P)}}\, \int d\eta \, \eta^{\triangle - n}\, (\xi_{n}\eta)^{\frac{n-d-1}{2}}\, K_{\frac{n-d-1}{2}}(2\xi_{n}\eta)\, T_{n}(\eta^{2}\, s_{ij}^{0}(\vec{\sigma})\, ),
\end{align}
where $\vec{\sigma}$ is the set of cross ratios that are invariant under the OP limit. 
\begin{align}
\vec{\sigma}\, :=\, \{\sigma_{1}, \dots, \sigma_{\frac{n(n-3)}{2}-1}\, \}
\end{align}
and, 
\begin{equation}
 \mathcal{N}_n = 2 \pi^{2}\, L^{-n\alpha(d) + d + 1}\, e^{(i\, 3\pi)}\, (2\pi)^{\frac{(n-1)}{2}} \left( \frac{L}{l_s}\right)^{\Delta-n+1} \left( \prod_{j=1}^{n} \frac{(-i)^{\Delta_j} C_{\Delta_j} }{\Gamma(\Delta_j)} \right)
\end{equation}
We now note that the 10 dimensional flat space amplitude $T_{n}$ can be written in terms of a dimensionless function ${\cal T}_{n}$ as 
\begin{align}
T_{n}^{D=10}\, =\, l_{s}^{4n-10}\, {\cal T}_{n}
\end{align}
However, since the bulk kinematics is in $D=5$, the effective 5 dimensional amplitude is related to $T_{n}$ as follows. For any $n$ point tree-level diagram with cubic vertices, we have the standard relation
\begin{align}
V - I = 1 
\end{align}
where $V$ is the number of cubic vertices and $I$ is the number of internal edges. As all our internal and external propagating modes are zero modes in $S^{5}$, each vertex contributes a factor of $V_{5}\, =\, \Omega_{5}\, L^{5}$, each propagator contributes a factor of $(Y_{0}\, Y_{0}^{\star})\, = V_{5}^{-1}\, =\, \frac{1}{\Omega_{5}}\, \frac{1}{L^{5}}$, and each external state contributes a factor of $\frac{1}{\sqrt{V_{5}}}\, \sim\, \frac{1}{\sqrt{\Omega_{5}}}\, \frac{1}{L^{\frac{5}{2}}}$. Hence combining all these factors, we can write the flat space amplitude with kinematics in 5 non-compact dimensions in terms of ${\cal T}_{n}$ as, 
\begin{align}\label{tnd5caltn}
T_{n}\, =\, \Omega_{5}^{(1-\frac{n}{2})}\, L^{5(1-\frac{n}{2})}\, l_{s}^{4n-10}\, {\cal T}_{n}
\end{align}
Substituting eqn.(\ref{tnd5caltn}) in eqn.(\ref{fin_npt0}) we get,  
\begin{align}\label{fin_npt}
{\cal A}_{n}(\xi_{n}, \vec{\sigma})\, \approx\, \frac{\mathcal{N}^{\prime}_n}{\sqrt{\textrm{det}^{\prime}(P)}}\, \int d\eta \, \eta^{\triangle - n}\, (\xi_{n}\eta)^{\frac{n-d-1}{2}}\, K_{\frac{n-d-1}{2}}(2\xi_{n}\eta)\, {\cal T}_{n}(\eta^{2}\, s_{ij}^{0}(\vec{\sigma})\, ),
\end{align}
where
\begin{align}\label{eq:norm_constant_an}
{\cal N}^{\prime}_{n}\, =\, (\, \frac{L}{l_s}\, )^{\triangle - 5n + 11}\, \Omega_{5}^{(1-\frac{n}{2})}\, 2 \pi^{2}\, e^{(i\, 3\pi)}\, (2\pi)^{\frac{(n-1)}{2}}\, \left( \prod_{j=1}^{n} \frac{(-i)^{\Delta_j} C_{\Delta_j} }{\Gamma(\Delta_j)} \right)
\end{align}
We can finally take the double scaling limit and define a boundary correlator in ${\cal N} = 4$ SYM theory as, 
\begin{align}\label{eq:anflat_final}
{\cal A}_{n}^{\textrm{flat}}(\xi_{n}, \vec{\sigma})\, =\, \lim_{\frac{L}{l_{s}}\, \rightarrow\, \infty}\, (\frac{l_s}{L})^{\triangle - 5n + 11}\, {\cal A}_{n}(\xi_{n}, \vec{\sigma})
\end{align}
This is the generalization of the result derived by Okuda and Penedones to generic $n$ point amplitudes in $d\, >\, n-3$ dimensions. For $d=4$ then the result is valid up to $n=6$. It can also be immediately verified that in the Four point case the RHS of eqn.(\ref{fin_npt}) matches with the result derived in \cite{Okuda:2010ym}  upto an overall irrelevant phase.

For $n > 4$ there is no canonical choice of the reduced correlator ${\cal F}_{n}$. However, one possibility is the following : 
\begin{align}
{\cal F}_{n}(\xi_{n}, \vec{\sigma})\, =\, \frac{{\cal A}_{n}^{\textrm{flat}}(g_{YM},\, \xi_{n}, \vec{\sigma})}{\prod_{i<j} P_{ij}^{m_{ij}}}
\end{align}
where the exponents $m_{ij}$ satisfy,
\begin{align}
\sum_{j \neq\, i} m_{ij}\, =\, \triangle_{i}
\end{align}
This ensures that ${\cal F}_{n}$ is invariant under projective re-scalings of $\{P_{1}, \dots, P_{n}\}$.\footnote{In \cite{Okuda:2010ym}, a specific choice of co-ordinates for $P_{1}, \dots P_{4}$ on the conformal boundary in the Four point case leads to the identity $\prod_{i<j} P_{ij}^{-m_{ij}}\, =\, \sigma^{- \triangle_{1} - \triangle_{2}}$ for $\triangle_{2} = \triangle_{1}$ and $\triangle_{4} = \triangle_{3}$.}
We end this section with a few remarks.
\begin{itemize}
    
 \item The OP limit has been shown to exist only for tree-level amplitudes. This is because, in the case of loop corrections, even if external states are localized on $\vec{l} = 0$ mode on $S^{5}$, the non-trivial KK modes do flow along the loops and hence the existence of OP limit becomes more subtle and its potential generalization is outside the scope of this paper.
\item An unresolved subtlety for the $n > 4$ point amplitude is to write the factor $\textrm{det}^{\prime}(P)$ in terms of invariant cross ratios. As $\sum_{i=1}^{n-1} \lambda_{i} = 0$ in the OP limit, we need to parameterize $\lambda_{i}\vert_{i=1}^{n-1}$ in terms of $\vec{\sigma}$ that satisfy this condition.\\
In the case of $n\, =\, 4$, OP chooses the following parameterization,
\begin{align}
\lambda_{1}\, =\, 8, \quad \lambda_{2} = -8 \sigma, \quad \lambda_{3}\, =\, -8 (1 - \sigma).
\end{align}
In this paper, we do not choose any specific parameterization of $\lambda_{i}\vert_{i=1}^{n-1}$ for $ n > 4$ in terms of $\vec{\sigma}$. It would be interesting to find a suitable parametrization for generic configuration that could assist us in further analyzing the OP limit of the SYM correlator in more detail. 
\item For any $n$ point amplitude (so long as $ D\,> n-2$) the OP limit of the dual correlator is a one dimensional integral over a scale variable $\eta$ of the flat space amplitude. More in detail, the integrand in the OP limit has the kinematics split in terms of the ``scale and angles'' as follows. We recall that the flat space momenta are given by, 
\begin{align}
k_{j} = - 2 \frac{\beta_{j}\, P_{j}}{l_{s}}\, \approx\, - 2 \eta \frac{P_{j}}{l_{s}} =:\, \eta \, \hat{k}_{j}
\end{align}
where $\hat{k}_{j}$ is a dimensionless Four vector. Thus, any flat space Mandelstam invariant can be written as 
\begin{align}
k_{i} \cdot k_{j} = \eta^{2} \hat{k}_{i} \cdot \hat{k}_{j} \, .
\end{align}
Inspired by \cite{Brown:2011pj}, we refer to this decomposition (which is enforced on us by the OP scaling) as the ``scale and angle'' decomposition of the kinematical variables. The soft limit of the flat space amplitude is then obtained by taking $\hat{k}_{i} \cdot \hat{k}_{n}\, \rightarrow\, 0$ inside the integral. In the case of tree-level dilaton bulk amplitude, the leading order soft theorem then takes the form \cite{DiVecchia:2015jaq}
\begin{align}
T_{n}\, =\, \left( 2- \sum_{i=1}^{n-1}\, \hat{k}_{i} \partial_{\hat{k}_{i}} \right)\, T_{n-1}(\eta^{2}\, {s}^0_{ij}\vert_{j < n})\, +\, O(\vert \hat{k}_{n} \vert) \, .
\end{align} 
\item In the strong coupling limit, the SYM correlation functions are dual to super-gravity amplitudes in flat space. In the double scaling (OP) limit we obtain a class of correlation functions which are dual to perturbative string amplitudes. This implies that even single trace unprotected operators (such as Konishi operators) whose conformal dimension scale as $\lambda^{\frac{1}{4}}$ do not decouple in the OP limit. This can be heuristically argued as follows. The dominant contribution of an operator of dimension $\triangle$  to the conformal block in the Lorentzian kinematics behaves as $e^{i \triangle\, \rho}\, \sim_{OP}\, O(1)$. That is by adjusting the Lorentzian cross ratio $\rho$ to scale precisely as $\lambda^{-\frac{1}{4}}$, the OP limit ensures that the boundary correlators are gauge theory duals of perturbative string amplitudes and not simply super-gravity amplitudes. 
\item In Sec.~\ref{soft_5pt} we prove that the soft factorization theorem stated above leads to a ``factorization theorem'' for tree-level correlators in the OP limit. These relations thus provide an interesting class of bootstrap constraints for Carrollian correlators that are dual to tree-level string amplitudes in flat space. 
\end{itemize}
\section{Double scaling limit in the presence of Gram constraints.}\label{gramimpose}
In this section we will extend the previous derivation to the case of interest, $D=5$. When space-time dimension is fixed, the space of Mandelstam invariants generated by $n$ momenta (which satisfy momentum conservation) is a variety inside the vector space generated by $\{s_{ij}\}$. This variety is obtained by solving the so called Gram constraints that arise due to linear dependence of $n-1$ vectors in $D \leq\, n-2$ dimensions. A simple dimension counting shows that the number of Gram constraints in 5 dimensions are $\frac{(n-6)(n-5)}{2}$. 
This entire set of constraints can be defined as follows. Let us assume that we have used momentum conservation to solve for $p_{n}$.  Let $M$ be the $n-1\, \times\, n-1$ matrix with
\begin{align}
M_{ij}\, =\, s_{ij}\, \forall\, 1\, \leq\, i \leq\, j\, \leq\, n-1
\end{align}
Then the set of Gram constraints is the set of all $6\, \times\, 6$ minors of the matrix $M$. We denote a minor (spanned by $\{s_{i_{m} j_{m}}\}_{m=1}^{6}$) as ${\cal C}(i_{1}, \dots,\, i_{6})$. We then have
\begin{align}
{\cal C}_{i_{1}\, \dots\, i_{6}}\, :=\, \textrm{Det} \, {\cal C}(i_{1},\, \dots,\, i_{6})\, =\, 0\, \, \, \forall\, \, \{ i_{1},\, \dots,\, i_{6}\, \}\, \in\, \{\, 1,\, \dots,\, n-1\, \}.
\end{align}
In the case of massless particles, the constraints can also be parametrized in terms of re-scaled Mandelstam variables $\{s^{(0)}_{ij}\}$ since ${\cal G}$ is a homogeneous polynomial of degree 12 in the scale variable $\eta$. Inside the kinematic space spanned by independent $\{s^{(0)}_{ij}\}$ the Gram variety is then defined as
\begin{align}
{\cal G}\, :=\, \cap_{a=1}^{\frac{(n-5)(n-6)}{2}}\, {\cal C}_{i_{1}\, \dots\, i_{6}}.
\end{align}
Before projecting onto the Gram variety, the set $\{s^{(0)}_{ij}\}$ is in bijection with the set of all cross ratios $\vec{\sigma}$ (for an explicit illustration of such a bijection, we refer the reader to the appendix \ref{app_mandelstam}). Thus for $D \leq\, n-2$, the set of physical cross ratios also belongs to the variety defined as
\begin{align}
{\cal G}_{\textrm{conf}}\, :=\, {\cal G}(\, \{s^{(0)}_{ij}(\vec{\sigma})\}\, ).
\end{align}
For an extensive discussion and seminal developments in the analysis of ${\cal G}_{\textrm{conf}}$ we refer the reader to \cite{Borovik:2026zgl}.

However, an explicit parameterization of the Gram variety is an extremely complicated computational algebraic geometry problem, and hence in this section we simply redo the entire derivation of the previous section by assuming that $\lambda_{0}$ is the eigenvalue tending to zero among the set of all non-trivial eigenvalues of the matrix $\{P_{ij}\}_{i,j=1}^{n}$. For concreteness, we only consider $n=7$.

The generalization of our derivation to arbitrary $n$ should be straightforward, although it will involve tedious linear algebra.

As seen in the previous section, for $D=5$, eqn.(\ref{Anflat1}) is valid for $\forall\, n$. For the convenience of the reader, we rewrite the equation below. 
For brevity, we suppress all the general factors that are not relevant to the present discussion. To leading order in $\frac{1}{L}$ we get 
\begin{align}\label{a7flat5d}
{\cal A}_{7}&\approx\, \frac{1}{L^{\frac{11}{2}}}\, \left(\frac{L}{l_s}\right)^{\Delta}\, \int_{0}^{\infty} \frac{d y}{y^{\frac{d+2}{2}}} \, e^{iy} \nonumber \\
& \times \, \prod_{i=1}^{7} \int_{0}^{\infty} d \beta_{i}\, \beta_{i}^{\triangle_{i}-1} e^{ \frac{i}{2 y} \left(\frac{L}{l_{s}}\right)^{2} \sum_{i,j} \beta_{i} \beta_{j} P_{ij}} \, i \, T_{7}\left(k_{7}= - \sum_{i=1}^{6} k_i\right)
\end{align}
In this case, the matrix $P_{ij}$ has co-rank 1 and hence has one zero eigen-value. Let the corresponding eigen-vector be denoted by $\vec{\Psi}_{-1}$. Thus, the spectrum of the matrix is of the form
\begin{align}
\textrm{Spec}(P_{ij})\, =\, \{\, 0,\, \lambda_{0}, \lambda_{1}, \dots,\, \lambda_{5}\, \}
\end{align}
with the corresponding eigen-vectors $\{\vec{\Psi}_{-1},\, \Psi_{0},\, \vec{\Psi}_{a}\vert_{a=1}^{5}\, \}$.
We also parameterize $\vec{\beta}$ as 
\begin{align}
\vec{\beta}\, =\, \nu_{-1}\, \vec{\Psi}_{-1}\, +\, \eta\, \vec{\Psi}_{0}\, +\, \sum_{a=1}^{5}\, \nu_{a}\, \vec{\psi}_{a}
\end{align}
Hence,
\begin{align}
\sum_{i,j} \beta_{i} \beta_{j} P_{ij}\, =\, \lambda_{0} \eta^{2} + \sum_{a=1}^{5}\, \lambda_{a} \nu_{a}^{2}\nonumber\\
d^{7}\vec{\beta}\, =\, d\eta\, d\nu_{-1}\, \prod_{a=1}^{5}\, d \nu_{a}
\end{align}
Similarly, in the presence of $\vec{\psi}_{-1}$, eqn.(\ref{etavnui}) is modified as follows. Let $\vec{\nu}\, =\, (\nu_{1}, \dots,\, \nu_{5})^{T}$. 
\begin{align}\label{etavnuim}
\int \prod_{i=1}^7 d \beta_{i} \beta_{i}^{\triangle_{i}-1}\, f(\vec{\beta}) =\, \int d\nu_{-1}\, d\eta \, d^{5}\vv{\nu}\ \prod_{i}((\ \eta + \nu_{-1} \Psi_{-1}^{i})\, + \sum_{a}\nu_{a}\, \Psi_{a}^{i})^{\triangle_{i} - 1} f(\eta, \vec{\nu})\nonumber\\
\hspace*{-0.4in}\overset{\textrm{OP}}{=}\, \int d\nu_{-1}\, d\eta \, d^{5}\vv{\nu}\ \left( \prod_{i}(\, \eta + \nu_{-1} \Psi_{-1}^{i}\, )^{\triangle_{i} - 1}\, f(\eta, \vec{\nu})\, +\, O(\nu_{a}\vert_{a=1}^{5})\, \right)
\end{align} 
To avoid notational clutter, we define
\begin{align}
\prod_{i=1}^{n}(\, \eta + \nu_{-1} \Psi_{-1}^{i}\, )^{\triangle_{i} - 1}\, =\,  \omega(\eta, \nu_{-1}, \vec{\triangle}).
\end{align}
We can now substitute eqn.(\ref{etavnuim}) in eqn.(\ref{a7flat5d}) and essentially follow the steps outlined in section(\ref{npt_gen}). The key differences between the two computations are summarized below. 
\begin{enumerate}
\item The Gaussian integral is over $d^{5}\vec{\nu}$ instead of $d^{6}\vec{\nu}$. 
\begin{align} 
\int d^{5}\vec{\nu}\, e^{\frac{i}{2y}\, (\frac{L}{l_{s}})^{2}\, \sum_{a=1}^{5}\, \lambda_{a}\, \nu_{a}^{2}}\, =\, \frac{1}{\sqrt{\prod_{a=1}^{5}\, \lambda_{a}}}\, (2\pi)^{\frac{5}{2}}\, e^{i\, \frac{5\pi}{4}}\, (\frac{l_{s}}{L})^{5}\, y^{\frac{5}{2}}
\end{align}
\item After integrating over $y$, we get $2 e^{i\frac{\pi}{4}}\, (\xi_{7}\, \eta)^{\frac{1}{2}}\, K_{\frac{1}{2}}(2 \xi_{7}\, \eta)$.
\end{enumerate}
Thus, finally, the right hand side of the OP limit is a two dimensional integral over $\nu_{-1},\, \eta$ which can be written as, 
\begin{align}\label{fin-7pt}
{\cal A}_{7}(\xi_{7},\, \vec{\sigma})\vert_{\lambda_{-1} = 0}\, &\approx\nonumber\\
&\hspace*{-0.7in}{\cal N}_{7}\,  \frac{1}{\sqrt{\prod_{a=1}^{5}\lambda_{a}}}\, \int\, d\eta\, \int d\nu_{-1}\, \omega(\eta, \nu_{-1},\, \vec{\triangle})\, (\xi_{7}\, \eta)^{\frac{1}{2}}\, K_{\frac{1}{2}}(2 \xi_{7}\, \eta)\, T_{7}(\{ s_{ij} = \frac{1}{l_{s}^{2}}\, \beta_{i}\, \beta_{j}\, P_{ij}\, \}\, )\vert_{{\cal G}}
\end{align}
\begin{align}
{\cal N}_{7}\, =\, 2 \pi^{2}\, L^{-\frac{11}{2}}\,  e^{3i\, \pi} \left( (2\pi)^{\frac{5}{2}}\, \left( \frac{L}{l_s}\right)^{\Delta-5}\,  \prod_{j=1}^{7} \frac{(-i)^{\Delta_j} C_{\Delta_j} }{\Gamma(\Delta_j)} \right)
\end{align}
We now note that the asymptotic behavior of the kernel $K_{\frac{1}{2}}$ will depend on whether $\xi_{7}$  is real or imaginary. And this in turn depends on the sign of $\lambda_{0}$ in eqn.\eqref{eq:xin2_def}.\footnote{For $n = 4$, $\xi_{4}^{2} > 0$ since $0 <\, \sigma\, <\, 1$ and $\rho^{2} < 0$ in the Lorentzian branch, however, at higher points determining the sign of $\xi^{2}$ requires a detailed analysis of Lorentzian kinematics,} However, for either sign the integral will be convergent. Since
\begin{align}
K_{\frac{1}{2}}(2\xi_{7}\eta)\, &=\, \sqrt{\frac{\pi}{4 \xi_{7} \eta}}\, e^{-2 \xi_{7}\, \eta}\, \quad \textrm{if}\, ~ \xi_{7}\, > 0\nonumber\\
K_{\frac{1}{2}}(2\xi_{7}\eta)\, &=\, \sqrt{\frac{\pi}{4 \kappa  \eta}}\, e^{- i (2 \kappa \eta + \pi/4 - i\epsilon \eta)}\, \quad \textrm{if} \, ~ \xi_{7}^{2}\, <\, 0
\end{align}
In the above equation, we choose the positive square root of $\xi_{7}^{2} = -\kappa^2$ in the first line. $i\epsilon$ in the second line ensures the convergence of the integral  as $\eta\, \rightarrow\, \infty$.

We can now replicate the steps of the previous section to obtain a map between the flat space amplitude ${\cal T}_{7}^{D=5}$ and the boundary correlators. Using  eqn.(\ref{tnd5caltn}) we find 
\begin{align}
T_{7}\, =\, \Omega^{-\frac{5}{2}}_5 \, l_{s}^{\frac{11}{2}}\, (\frac{L}{l_{s}})^{-\frac{25}{2}}\, {\cal T}_{7}.
\end{align}
Hence,
\begin{align}
{\cal A}_{7}(\xi_{7},\, \vec{\sigma})\vert_{\lambda_{-1} = 0}\, &\approx\nonumber\\
&\hspace*{-0.7in}{\cal N}^{\prime}_{7}\,  \frac{1}{\sqrt{\prod_{a=1}^{5}\lambda_{a}}}\, \int\, d\eta\, \int d\nu_{-1}\, \omega(\eta, \nu_{-1},\, \vec{\triangle})\, (\xi_{7}\, \eta)^{\frac{1}{2}}\, K_{\frac{1}{2}}(2 \xi_{7}\, \eta)\, {\cal T}_{7}(\{ s_{ij} = \frac{1}{l_{s}^{2}}\, \beta_{i}\, \beta_{j}\, P_{ij}\, \}\, )\vert_{{\cal G}}
\end{align}
with, 
\begin{align}
{\cal N}_{7}^{\prime}\, =\, 2 \pi^{2}\, \Omega^{-\frac{5}{2}}_5 \, \left( \frac{L}{l_s} \right)^{-18}\, e^{3i\, \pi} \left( (2\pi)^{\frac{5}{2}}\, \left( \frac{L}{l_s}\right)^{\Delta-5}\,  \prod_{j=1}^{7} \frac{(-i)^{\Delta_j} C_{\Delta_j} }{\Gamma(\Delta_j)} \right)
\end{align}
So finally, for half BPS operators with $\Delta_i=4$
\begin{align}
{\cal A}_{7}^{\textrm{flat}}(\xi_{7}, \vec{\sigma})\vert_{\lambda_{-1}\, =\, 0}\, =\, \lim_{\frac{L}{l_{s}}\, \rightarrow\, \infty}\, \left( \frac{l_s}{L} \right)^{5}\,  {\cal A}_{7}(\xi_{7}, \vec{\sigma})\vert_{\lambda_{-1}\, =\, 0}
\end{align}
This is the main result of the present section. It shows that the flat space limit of tree-level bosonic string amplitude in $\mathbb{R}^{1,9}$ with kinematics in a co-dimension 5 hyper-plane is dual to the double scaling limit of  7 point correlator of $\frac{1}{2}$ BPS scalars in ${\cal N} = 4$ SYM theory which is defined on a specific variety inside the space of cross-ratios.

\medskip

Several comments are in order at this stage:
\begin{itemize}
\item $\lambda_{-1} = 0$ denotes the fact that $\vec{\Psi}_{-1}$ is a null Eigen-vector of the matrix $P_{ij}$ before taking the OP limit. Hence, LHS is a conformal correlator evaluated on one branch of the ${\cal G}_{\textrm{conf}} = 0$ variety. This implies that on the right hand side, we have evaluated ${\cal T}_{7}$ on the same branch of ${\cal G}$ since $\textrm{det}^{\prime}(P) = 0\, \implies\, \textrm{det}(s_{ij})\, =\, 0$. 
\item  The core difference between eqn.(\ref{fin_npt}) and eqn.(\ref{fin-7pt}) is the following. Apart from exponents of $\frac{L}{l_{s}}$, the key structural difference between the two equations is that in dimensions $D = 5$ the OP limit generates a two dimensional integral transform of the flat space amplitude.
\end{itemize}
So far we have not specified the integration range of $\nu_{-1}$. Unlike the scale variable $\eta\, \in\, [0,\, \infty)$, we need to determine the domain of $\nu_{-1}$. We now prove that this domain is compact.

\medskip

In the OP limit
\begin{align}
\beta_{i}\, =\, \nu_{-1}\, \Psi_{-1 (i)} + \eta\, \Psi_{0\, (i)} .
\end{align}
Since each Schwinger parameter is positive, we have the following.
\begin{align}
\eta\, \Psi_{0\, i}\, +\, \nu_{-1}\, \Psi_{\perp(i)}\, \geq\, 0
\end{align}
As we have chosen $\vec{\Psi}_{0}\, = \frac{1}{\sqrt{7}}\, (1,\, \dots,\, 1)^{T}$.  it is clear that the positivity of Schwinger parameter $\beta_{i}$ corresponds to, 
\begin{align}\label{numinus1pm}
\frac{\eta}{\sqrt{7}} + \nu_{i}\, \Psi_{-1 i}\, >\, 0
\end{align}
Now since 
\begin{align}
\vec{\Psi}_{0}\, \cdot \vec{\Psi}_{-1} = 0,
\end{align}
we see that at least some components of $\Psi_{-1}$ have to be negative. Eqn. (\ref{numinus1pm}) then implies that for a fixed $\eta$, $\nu_{-1}$ belongs to a convex region within $\mathbb{R}^{2}_{+}$. This can be seen as follows. Let us assume that 
\begin{align}
\Psi_{\perp j} \geq\, 0\, \quad \forall\, 1 \leq\, j \leq\, J\nonumber\\
\Psi_{\perp j} \leq\, 0\, \quad \forall\, J\, <\, j\, \leq\, n
\end{align}
Then 
\begin{align}
\nu_{-1} > - \frac{\eta}{\sqrt{7}\, \Psi_{\perp j}} \, \quad \forall\, 1\, \leq\, j\, \leq J\nonumber\\
\nu_{-1} < \frac{\eta}{\sqrt{7} \,|\Psi_{\perp j}|} \, \quad \forall\, J+1 < j \leq\, n
\end{align}
The convergence of the integral over $\nu_{-1}$ is now at fixed $\eta$ is obvious, and the final integral over $\eta$ is convergent thanks to the asymptotic properties of $K_{\frac{1}{2}}(z)$, for $z > 0 \in \mathbb{R}$ or $z^{2} <  0$.
%%%%%%%%%%%%%%%%%%%%%%%%%%%%%%%%%
%%%%%%%%%%%%%%%%%%%%%%%%%%%%%%%%%%%%%%%%%%%%%%%%%%%%%%%%
%%%%%%%%%%%%%%%%%%%%%%%%%%%%%%%%%%%%%%%%%%%%%%%%%%%%
\section{Mapping OP kinematics to Null infinity}
\label{null_map}
In~\cite{Okuda:2010ym}, Okuda and Penedones proved that the OP limit for the 4-point amplitude is realized by taking the boundary insertions $P_i$ to approach the future and past light cone of a common reference point $x_{0}$ at a rate governed by the strong coupling limit. They explicitly constructed a specific configuration of 4 boundary points that  satisfied the equations \eqref{pijlambda0} and \eqref{lambda0largeR}, with the deviation of each $P_i$ from the light cone scaling as $\mathcal{O}(N^{-1/4})$. This result was stated in terms of the embedding-space coordinates $P_i$, without specifying how this deviation is achieved in any particular set of boundary coordinates.

In this section, we discuss this realization explicitly. In particular, we take a limit of external insertions such that two of the boundary insertions  approach $\tau = -\frac{\pi}{2}$ and the remaining two approach $\tau = \frac{\pi}{2}$ hypersurfaces at the following rate
\begin{align}
\tau_{i} = \pm \frac{\pi}{2} + \mathcal{O}\left(\frac{1}{N^{1/4}}\right) \, .
\end{align}
Moreover, we show that this specific limit can be used to recast the correlators of ${\cal N} = 4$ theory in infinite coupling limit as correlators of some (hitherto unknown) Carrollian QFT.

We are interested in the behaviour of $\rho$ as $N \to \infty$. $\rho$ is given by
\begin{equation}\label{rho}
\begin{gathered}
\sinh^2\rho = \frac{\det{P_{i}\cdot P_j}}{4 (P_1\cdot P_3) (P_2\cdot P_4) (P_1\cdot P_2) (P_3\cdot P_4)}\\
= \frac{\left( 1 + \frac{(P_1\cdot P_3) (P_2\cdot P_4)}{(P_{1} \cdot P_2) (P_{3} \cdot P_4)} - \frac{(P_1\cdot P_4)(P_2\cdot P_3)}{(P_{1} \cdot P_2) (P_{3} \cdot P_4)}\right)^2}{ 4\frac{(P_1\cdot P_3) (P_2\cdot P_4)}{ (P_1\cdot P_2) (P_3\cdot P_4)}} - 1
\end{gathered}
\end{equation}

The OP limit requires $\rho \sim \mathcal{O}\left( \frac{1}{N^{1/4}} \right)$.
In terms of global coordinates, $ P_{i}\cdot P_j $ is given by
\begin{equation}\label{Pij_global}
P_{i}\cdot P_j = -\cos(\tau_i-\tau_j) +  \vv{e}_i \cdot \vv{e}_j
\end{equation}
Let us take the value of the global times $ \tau_i $ corresponding to the four scattering states as 
\begin{equation}\label{in_fin_patch1}
 \tau_{1,2} = -\frac{\pi}{2} + \frac{\tilde{u}_{1,2}}{L}, \ \ \tau_{3,4} = \frac{\pi}{2} + \frac{\tilde{u}_{3,4}}{L}
\end{equation}
$\tilde{u}$ has dimension of length. However, instead of $\tilde{u}$ we will work with a dimensionless combination $u=\frac{\tilde{u}}{l_s} $. In other words, $u$  is measured in units of string length. In terms of $u$ \eqref{in_fin_patch1}becomes
\begin{equation}\label{in_fin_patch}
 \tau_{1,2} = -\frac{\pi}{2} + \frac{u_{1,2}}{L/l_s}, \qquad \tau_{3,4} = \frac{\pi}{2} + \frac{u_{3,4}}{L/l_s} \, .
\end{equation}
We also introduce a short hand notation:
\begin{equation}
 u_{ij} = u_i-u_j, \, {\bf e}_{ij} = \vv{e}_i \cdot \vv{e}_j.\, 
\end{equation}
We now substitute \eqref{in_fin_patch} in \eqref{rho} and expand in the $L/l_s \to \infty$ limit. Since we are working with the boundary quantities, it is more natural to phrase this as the limit $N\to \infty$ rather than the limit $L/l_s \to \infty$. The expansion is given by
\begin{equation}\label{rho_exp}
\begin{gathered}
\sinh^2 \rho = \frac{\left( 1 + \frac{(1+{\bf e}_{13}) (1+{\bf e}_{24})}{(1-{\bf e}_{12}) (1 - {\bf e}_{34})} - \frac{(1+{\bf e}_{14})(1+{\bf e}_{23})}{(1-{\bf e}_{12}) (1-{\bf e}_{34})}\right)^2}{ 4\frac{(1+{\bf e}_{13}) (1+{\bf e}_{24})}{(1-{\bf e}_{12}) (1 - {\bf e}_{34})}} - 1 + \mathcal{O}\left( \frac{1}{\sqrt{N}} \right) \, .
\end{gathered}
\end{equation}
The OP limit requires 
\begin{equation}\label{OP_cond}
    \rho^2 \sim \mathcal{O}\left( \frac{1}{\sqrt{N}}\right)\,.
\end{equation} 
In particular, this means that no term of $ \mathcal{O}\left( \frac{1}{N^{1/4}}\right) $ can be present in the expansion of $ \sinh^2\rho $. As equation \eqref{rho_exp} shows, this is indeed the case for our configuration \eqref{in_fin_patch}: $ \mathcal{O}\left( \frac{1}{N^{1/4}}\right) $ the term is absent. This is not a generic feature of an arbitrary configuration of insertions approaching the light cone - it is a direct consequence of our choice $\tau_{1,2,3,4} \to \pm \frac{\pi}{2}$ as $N \to \infty$. In this sense, although we started with this configuration as an ansatz, the OP limit effectively forces this choice. More precisely, fixing the initial patch at $ -\frac{\pi}{2} $ forces the final patch to be at $ \frac{\pi}{2} $.

However, the configuration \eqref{in_fin_patch} does not automatically set the $\mathcal{O}(1)$ term to 0. We must therefore impose 
\begin{equation}\label{constr}
    \frac{\left( 1 + \frac{(1+{\bf e}_{13}) (1+{\bf e}_{24})}{(1-{\bf e}_{12}) (1 - {\bf e}_{34})} - \frac{(1+{\bf e}_{14})(1+{\bf e}_{23})}{(1-{\bf e}_{12}) (1-{\bf e}_{34})}\right)^2}{ 4\frac{(1+{\bf e}_{13}) (1+{\bf e}_{24})}{(1-{\bf e}_{12}) (1 - {\bf e}_{34})}} - 1 = 0 \, ,
\end{equation}
as a constraint on the directions of the four massless particles required for the OP condition \eqref{OP_cond} to hold. This constraint simplifies considerably for a special choice of the angular variables, which we now make explicit. Since the boundary is 4 dimensional, the unit vectors take the form
\begin{equation}
\begin{split}
\vv{ e}_i = \lbrace \sin \theta_i  \sin \phi_i \cos \psi_i ,\sin \theta_i  \sin \phi_i \sin \psi_i,\sin \theta_i  \cos \phi_i ,\cos \theta_i  \rbrace, \qquad \textnormal{for} \ i=1,2,\\
\vv{ e}_i = \lbrace \sin (\pi-\theta_i)  \sin (\pi-\phi_i) \cos (\pi+\psi_i) ,\sin (\pi-\theta_i)  \sin (\pi-\phi_i) \sin (\pi+\psi_i), \\
\sin (\pi-\theta_i)  \cos (\pi-\phi_i) ,\cos (\pi-\theta_i)  \rbrace, \qquad \textnormal{for} \ i=3,4.
\end{split}
\end{equation}
Let us consider that all the particles are on the equatorial sphere, that is,
\begin{equation}
\theta_i = \frac{\pi}{2}, \qquad i=1,\ldots,4.
\end{equation}
Introducing a complex variable, $ z_i = \tan\left( \frac{\phi_i}{2}\right)e^{i\psi_i} $, we obtain,
\begin{equation}
\begin{gathered}
\sinh^2 \rho = \frac{(z-\bar z)^2}{4 z \bar z} + \mathcal{O}\left( \frac{1}{\sqrt{N}} \right)
\end{gathered}
\end{equation}
with 
\begin{equation}
z = \frac{z_{13}z_{24}}{z_{12}z_{34}}, \qquad \bar z = \frac{\bar z_{13} \bar z_{24}}{\bar z_{12}\bar z_{34}} \, .
\end{equation}
On the equatorial sphere, the OP constraint \eqref{constr}  therefore reduces to the requirement that cross ratio $z$ must be real.

We now use the mapping of the local patch around $\tau = \pm\frac{\pi}{2}$ to $ {\I}^{+}$ to re-write $\xi_{4}$ in terms of Carrollian co-ordinates.
The definition of $ \xi_4 $ is given by,
\begin{equation}
\begin{gathered}
\xi_4^2 = - \lim_{\textnormal{det}P_{ij}\to 0, \frac{L}{l_s} \to \infty} \left(\frac{L}{l_s}\right)^2 \frac{\textnormal{det}P_{ij}}{4 P_{12}P_{34}\sqrt{P_{13} P_{24} P_{14} P_{23} }} \\
= - \lim_{\textnormal{det}P_{ij}\to 0, N \to \infty} \sqrt{ g_{YM}^2 N}  \sqrt{\frac{P_{13}  P_{24 }}{P_{14} P_{23} }} \left[ \frac{\left( 1 + \frac{P_{13}  P_{24 }}{P_{12} P_{34} } - \frac{P_{14}  P_{23 }}{P_{12} P_{34} } \right)^2}{ 4\frac{P_{13}  P_{24 }}{P_{12} P_{34} }} - 1 \right] \, .
\end{gathered}
\end{equation}

Let us use some short hand notations:
\begin{equation}\label{fs}
\begin{gathered}
r^{1324}_{1234} = \frac{(1+{\bf e}_{13}) (1+{\bf e}_{24})}{(1-{\bf e}_{12}) (1 - {\bf e}_{34})} , \qquad r^{1423}_{1234} = \frac{(1+{\bf e}_{14})(1+{\bf e}_{23})}{(1-{\bf e}_{12}) (1-{\bf e}_{34})}, \qquad r^{1324}_{1423} = \frac{(1+{\bf e}_{13}) (1+{\bf e}_{24})}{(1+{\bf e}_{14}) (1 + {\bf e}_{23})}\\
f_{1234} = \frac{u_{12}^2}{(1-{\bf e}_{12})}+\frac{u_{34}^2}{(1-{\bf e}_{34})}, \qquad f_{1423} = \frac{u_{14}^2}{(1+{\bf e}_{14})} + \frac{u_{23}^2}{(1+{\bf e}_{23})},\\
 f_{1324} = \frac{u_{13}^2}{(1+{\bf e}_{13})} + \frac{u_{24}^2}{(1+{\bf e}_{24})}.
\end{gathered}
\end{equation}
Using the above notations we get,
\begin{equation}\label{xi4sq}
\xi^2_4 = \left(\xi_4^0\right)^2 + \mathcal{O}\left( \frac{1}{N^{1/2}} \right)
\end{equation}
where
\begin{equation}\label{xio42}
\begin{gathered}
 (\xi^0_4)^2  = - \frac{\sqrt{r^{1324}_{1423}} (1+r^{1324}_{1234}-r^{1423}_{1234})}{8 \,  r^{1324}_{1234} }\left[(f_{1324}-f_{1234})(1-r_{1234}^{1324}-r^{1423}_{1234}) +2(f_{1423}-f_{1234})r^{1423}_{1234} \right] \\
\end{gathered}
\end{equation}
The above expression can be simplified using the flat space momentum conservation in the OP limit. In section \ref{npt_gen}, we defined the bulk momenta as (equation \eqref{fsmom})
\begin{equation}
    k_i := - \frac{2\beta_i P_i}{l_s}.
\end{equation}
In the OP limit, $\beta_i \sim \eta$ for all $i$, and momentum conservation therefore implies
\begin{equation} \label{mcp}
    \sum_{i=1}^4 P_i = 0.
\end{equation}
Using the parameterization \eqref{P_param} for $P_i$ and \eqref{in_fin_patch} for $\tau_i$, and taking the limit $N \to \infty$, one can show that equation \eqref{mcp} reduces to
\begin{equation}\label{mce}
    \sum_{i=1}^4 \vec{e}_i = 0.
\end{equation}
Equation \eqref{mce} can be used to write ${\bf e}_{34} = {\bf e}_{12}$, ${\bf e}_{24}={\bf e}_{13}$, ${\bf e}_{23}={\bf e}_{14}$, $1+{\bf e}_{12}+{\bf e}_{13}+{\bf e}_{14}=0$. Using these, \eqref{xio42} dramatically simplifies to
\begin{equation}\label{xi40}
    \xi^0_4 = i\frac{(u_1 + u_2-u_3 -u_4)}{\sqrt{2(1-{\bf e}_{12})}} \, .
\end{equation}

%%%%%%%%%%%
We now have all the tools at our disposal to define a Carrollian correlator which is obtained via OP limit of the 4 point SYM correlator.

 Okuda-Penedones \cite{Okuda:2010ym} considered 4-dilaton scattering amplitude in type IIB 10D string theory and computed the boundary correlator in the $ N \to \infty $ limit. It is given by,
\begin{equation}\label{OP_plan_sym}
\mathcal{F}_{\textnormal{planar}} = i g_{YM}^4 \frac{\pi^2}{9}\frac{(1-\sigma+\sigma^2)^2}{(\sigma(1-\sigma))^{3/2}}\frac{\Gamma(14)}{(2\xi_4)^{15}} + \cdots
\end{equation}
where the ellipsis denote the higher power terms in $\frac{1}{\xi_4}$. In the $ N \to \infty $ limit, $ \sigma = \sqrt{r^{1324}_{1234}} + \mathcal{O}\left( \frac{1}{N^{1/2}} \right) $. Thus, the 4-point function in the strong coupling limit is given by,
\begin{equation}\label{4ptfuncsc}
\begin{gathered}
\mathcal{F}_{\textnormal{planar}} = i g_{YM}^4 \frac{\pi^2}{9 }\frac{(1-\sqrt{r^{1324}_{1234}}+r^{1324}_{1234})^2}{\left[ \sqrt{r^{1324}_{1234}}(1-\sqrt{r^{1324}_{1234}})\right]^{3/2}}\frac{\Gamma(14)}{(2\xi_4^0)^{15}} + \cdots \\
= - g_{YM}^4 \frac{\pi^2 \, \Gamma(14)}{2^{15/2} 9} \frac{\left[(1+{\bf e}_{13})^2 + (1+{\bf e}_{13})(1+{\bf e}_{14}) + (1+{\bf e}_{14})^2\right]^2}{(1-{\bf e}_{12})\left[(1+{\bf e}_{13})(1+{\bf e}_{14})\right]^{3/2}} \frac{(1-{\bf e}_{12})^{15/2}}{(u_1+u_2-u_3-u_4)^{15}} + \dots
\end{gathered}
\end{equation}
We close this section with a few remarks on the results above. As shown in \cite{Okuda:2010ym}, equation \eqref{OP_plan_sym} follows from first taking the small-$\alpha'$ expansion of the tree-level 4-dilaton scattering amplitude and then applying the OP integral transform. Thus, equation \eqref{OP_plan_sym} tells us that the small-$\alpha'$ expansion is equivalent to a large-$\xi_4$ expansion. The leading term in this large-$\xi_4$ expansion is precisely the SYM 4-point function in the supergravity approximation.

We then used the map \eqref{in_fin_patch} to re-express Four point correlator in terms of the coordinates on $\I^+ $. Here $\xi_4^0$ depends on the null coordinates $u_i, v_i$ together with the angular variables, and can be made large -- indeed taken to infinity -- at fixed angles simply by sending the separation between the null coordinates to infinity. The supergravity approximation is therefore realized, in our setup, by taking the boundary insertion points to be infinitely separated along the null directions. The bulk stringy effects can be probed by evaluating the Carrollian correlators at finite affine separation.\\

We have asserted that the OP limit can be used to define a set of Carrollian correlators on ${\I}^{+}$. A necessary condition for this assertion is that ${\cal F}_{\textrm{planar}}$ should satisfy the global Ward identities associated to 5 space-time translations. In other words, we expect that  
\begin{align}
\sum_{i=1}^{4} f_{I}(e_{i}) \partial_{u_{i}}\, {\cal F}_{\textrm{planar}}\, =\, 0\,~ \forall\, I\, \in\, \{0,\, \dots\, 4\} \, ,
\end{align}
where $f_{I}$ are the 5 generators of global translations of ${\I}^{+}$. Although the expression on RHS of eqn(\ref{4ptfuncsc}) is manifestly invariant only under global time-translations ($u_{i}\, \rightarrow\, u_{i} + a$), we will now argue that they satisfy all of the global Ward identities and as a result unambiguously qualify as Carrollian correlators.

It was proved in \cite{Okuda:2010ym} that ${\cal F}_{\textrm{planar}}$ is dual to the flat space S-matrix with emergent momentum conservation in the double-scaling limit. From the boundary perspective, the double scaling limit involves taking a limit where the external insertions approach a light cone of a common reference point.\footnote{The choice of this reference point is pure gauge and as a result, the Four point function in the OP limit is invariant under boundary space-time translations.} Thus, the double scaling limit generates light cone correlators which satisfy $d+1$ dimensional momentum conservation.

As the future (past) null infinity $\mathbb{R} \times S^{3}$ is a null manifold that admits a faithful representation of the space-time translation symmetry in $\mathbb{R}^{1,d}$, we have shown that the 4 point correlation functions in the OP limit can hence be mapped onto the correlators on ${\I}^{+}$ which satisfy the global Ward identities associated with space-time translations.

Although explicit mapping of doubly scaled correlators in $\partial$ AdS$_{5}$ to a set of Carrollian correlators has been done only for 4 point functions in this section, we believe that it can be generalized to higher point functions. For instance, in the case of the 5 point function, such a mapping would involve the choice of 4 of the 5 independent cross ratios, $\vec{\sigma}$ such that in the OP limit $\rho\, \rightarrow\, 0$, $\vec{\sigma}$ will only depend on the angular co-ordinates of the external insertions. As seen by explicit computation in the 4 point case, we believe that the remaining variable $\xi_{5}$ then will (at leading order in OP limit) be a function of $\{ u_{i}, {\bf e}_{i}\, \}_{i=1}^{5}$. However, an explicit expression of $\{\xi_{5}, \vec{\sigma}\}$ is beyond the scope of the present work.

%%%%%%%%%%%%%%%%%%%%%%%%%%%%%%%%%%%%%%%
\section{Soft limit of 5 point Correlator in the OP limit}
\label{soft_5pt}
In this section, we argue that the soft factorization of tree-level Dilaton amplitudes in $\mathbb{R}^{1,9}$ imply a specific factorization property for the SYM correlators in the OP limit. While this property appears highly non-trivial in a strongly coupled gauge theory, we will be primarily interested in recasting it as a constraint on the Carrollian theory that we are seeking.

\medskip

Let the integral transform which maps a flat space string amplitude to double scaled limit of SYM correlator be denoted as
\begin{align}
{\cal A}_{5}^{\textrm{flat}}(\xi_{5}, \vec{\sigma})\, =\, \frac{{\cal N}_{5}}{\sqrt{\textrm{det}^{\prime}(P)(\vec{\sigma})}}\, (\hat{I}_{5} {\cal T}_{5})(\xi_{5}, \vec{\sigma}).
\end{align}
In the above equation, $\hat{I}_{5}$ is the integral transform involving $K_{0}$ 
\begin{equation}
    \mathcal{A}_{5}^{\textrm{flat}}(\xi_5,\vec{\sigma}) = \frac{\mathcal{N}_5}{\sqrt{\det'(P)(\vec{\sigma})}}  \int d\eta \, \eta^{15}\, K_0(2 \xi_5 \eta) \, {\cal T}_5(\eta^2 s^0_{ij}(\vec{\sigma})) \, .
\end{equation}
This transform is invertible. More in detail, there exists $\hat{I}_{5}^{-1}$ such that
\begin{align}
{\cal T}_{5}(\eta^{2} s_{ij}^{0})\, =\, \hat{I}_{5}^{-1} (\, \sqrt{\textrm{det}^{\prime}(P)(\vec{\sigma})}\, {\cal A}^{\textrm{flat}}_{5}\, )
\end{align}
$\hat{I}_{5}^{-1}$ can be derived as follows. By using the identity, 
\begin{align}
\int_{0}^{\infty}\, d\xi_{5}\, \xi_{5}^{\alpha - 1}K_{0}(2 \xi_{5} \eta)\, =\, \frac{1}{4} \Gamma\left(\frac{\alpha}{2}\right)^{2} \eta^{-\alpha}
\end{align}
we get, 
\begin{align}
\int_{0}^{\infty} d \xi_{5} \xi_{5}^{\alpha - 1} {\cal A}^{\textrm{flat}}_{5}(\xi_{5}, \vec{\sigma})\, \sqrt{\textrm{det}^{\prime}(P)(\vec{\sigma})} =\, \frac{{\cal N}_{5}}{4}\, \Gamma\left(\frac{\alpha}{2}\right)^{2} \int d\eta\, \eta^{15 - \alpha}\, {\cal T}_{5}(\eta^{2} s_{ij}^{0}).
\end{align}
The map from ${\cal A}^{\textrm{flat}}_{5}(\xi_{5}, \vec{\sigma})$ to the flat space amplitude with scale  variable $\eta$ and angle variables $\{s^{0}_{ij}\}$ can then be derived using inverse Mellin transform. 
\begin{align}\label{t5fromf5}
{\cal T}_{5}(\eta^{2}\, s_{ij}^{0})\, =\, \frac{4}{{\cal N}_{5}}\, \int_{-i\infty}^{i\infty} \frac{d\alpha}{2\pi i}\, \frac{1}{\Gamma(\frac{\alpha}{2})^{2}}\, \eta^{\alpha-16}\, \int_{0}^{\infty} d \xi_{5} \xi_{5}^{\alpha - 1} {\cal F}_{5}(\xi_{5}, \vec{\sigma})\, \sqrt{\textrm{det}^{\prime}(P)(\vec{\sigma})}\nonumber\\
=:\, \hat{I}_{5}^{-1} (\, {\cal A}^{\textrm{flat}}_{5} \sqrt{\textrm{det}^{\prime}(P)(\vec{\sigma})}\, )
\end{align}
We can now use the sub-leading soft dilaton theorem (see e.g. \cite{DiVecchia:2015jaq}) for tree-level string amplitude on the LHS of eqn(\ref{t5fromf5}) which leads to the following recursion relation between 5 point and 4 point correlation functions,
\begin{align}
\lim_{\vert k_{5} \vert\, \rightarrow\, 0}\, \hat{I}_{5}^{-1} \left( \, {\cal A}^{\textrm{flat}}_{5} \sqrt{\textrm{det}^{\prime}(P)(\vec{\sigma})}\, \right)\, =\, (2\, -\, \sum_{i=1}^{4}k_{i}\partial_{k_{i}}\, )\, {\hat I}_{4}^{-1} \left(\, {\cal A}^{\textrm{flat}}_{4}\, \sqrt{\textrm{det}^{\prime}(P_{4})}\, \right)
\end{align}
where $k_{5}^{\mu}$ is the momentum of the soft dilaton.

\medskip

It is important to note that unlike the sub-leading soft graviton theorem, in the soft limit the 5 point amplitude does not depend on the direction of the soft dilaton in the sense that
\begin{align}
\nabla_{\hat{n}} \lim_{\vert k_{5} \vert\, \rightarrow\, 0}\, {\cal T}_{5}\, =\, \nabla_{\hat{n}} \lim_{\vert k_{5} \vert\, \rightarrow\, 0}\, \hat{I}_{5}^{-1} \left( \, {\cal A}^{\textrm{flat}}_{5} \sqrt{\textrm{det}^{\prime}(P)(\vec{\sigma})}\, \right)\, =\, 0
\end{align}
where $\hat{n}^{\mu} = \frac{1}{\vert k_{5} \vert}\, k^{\mu}$. 

\medskip

Of the 4 cross ratios $\vec{\sigma}$ we will choose $\sigma_{2}$ to be independent of $P_{5}$,
\begin{align}
\sigma_{2}\, =\, \sigma
\end{align}
where $\sigma\, \sim\,  \frac{(s^{(0)}_{23})^2}{s^{(0)}_{12}}$ (this $\sigma$ is different from eq.\eqref{eq:sigma_rho_def} and this change is merely due to calculational convenience) is the cross ratio for 4 point correlator. Hence, $\sigma_{1},\, \sigma_{3}, \sigma_{4}$ depend on $P_{5}$. We thus find the identity
\begin{align}\label{i5inversesoft}
\lim_{\vert k_{5}\vert\, \rightarrow\, 0}\, \hat{I}_{5}^{-1} \left(\, {\cal A}^{\textrm{flat}}_{5} \sqrt{\textrm{det}^{\prime}(P)(\vec{\sigma})}\, \right)\, =\, \left( 2- \sum_{i=1}^{4}k_{i}\partial_{k_{i}} \right)\, \hat{I}_{4}^{-1} \left(\, {\cal A}^{\textrm{flat}}_{4}\, \sqrt{\textrm{det}^{\prime}(P_{4})}\, \right)\, 
\end{align} 
where, $\hat{I}_{4}^{-1}$ is the inverse Laplace transform defined in \cite{Okuda:2010ym}. The LHS of eqn.(\ref{i5inversesoft}) is (the inverse transform) of 5 point correlator in the double scaling limit in which the 4 cross ratios $\vec{\sigma}$ lie on the variety defined in eqn.(\ref{softvar}) while $\rho_{5}$ is kept fixed. (See the argument below eqn.(\ref{i5soft}).) On the other hand RHS only depends on $\xi_{4},\, \sigma_{2}$. This simply reflects the fact that in the soft limit, the 5 point amplitude is independent of $k_{5}^{\mu}$.

\medskip

Let us define the inverse kernel $\hat{I}_{5}^{soft}$ in the soft limit as follows.
\begin{align}\label{i5soft}
(\hat{I}_{5}^{soft}f)(\xi_{5}^{soft},\, \vec{\sigma}_{i}^{soft})\, =\, \int_{0}^{\infty}\, \eta^{15}\, K_{0}(2\xi_{5}^{soft}\, \eta)f(\eta, \{s^{0}_{ij}\})\vert_{s^{0}_{i5}\, \rightarrow\, 0}
\end{align}
where $\vec{\sigma}^{\textrm{soft}}$ is a point in co-dimension one variety defined by eqn.(\ref{softvar}). In the soft limit $\xi_{5}\, \rightarrow\, \xi_{5}^{soft}$. This can be seen by explicitly evaluating $\xi_{5}$ in the soft limit in terms of the conformally invariant distance $\rho$. 
\begin{align}
\rho_{5}\, =\, \frac{\textrm{det}(P_{ij})}{P_{12}\, \dots\, P_{51}}
\end{align}
\begin{equation}
    \xi^2_{5}\, =\, -\, \frac{1}{2}\, \lim_{\frac{L}{l_{s}}\, \rightarrow\, \infty}\, \left(\frac{L}{l_{s}}\right)^{2}\, \lambda_{0} \, .
\end{equation}
And since, 
\begin{align}
\rho_{5}\, =\, \lambda_{0}\frac{\textrm{det}^{\prime}(P_{ij})}{P_{12}\, \dots\, P_{51}} \, .
\end{align}
We see that
\begin{align}
\xi^2_{5}\, =\, -\frac{1}{2}\, \lim_{OP}\, \bigg(\, \frac{(P_{12}\, \dots\, P_{51})}{\textrm{det}^{\prime}(P_{ij})}\, \left(\frac{L}{l_{s}}\right)^{2}\, \rho_{5}\, \bigg) \, ,
\end{align}
where $\lim_{OP}$ denotes the limit in which $P_{1}, \dots P_{5}$ satisfy,
\begin{align}
P_{i} \cdot P_{0} \sim\, \frac{1}{L} \, ,
\end{align}
for the reference null vector $P_{0}$.\footnote{We can of course re-write these limits in terms of $x_{ij}^{2} = - 2 P_{i} \cdot P_{j}$.}
We can write $\xi_{5}$ in a more compact notation as, 
\begin{equation}\label{xi5forsoftlimit}
    \xi^2_{5}\, =\, -\, \frac{1}{2}\, {\cal E}(\sigma_{1}, \dots\, \sigma_{4})\, \lim_{L\, \rightarrow\, \infty} \left(\frac{L}{l_{s}}\right)^{2} \rho_{5} \, ,
\end{equation}
where
\begin{align}
{\cal E} :=\, \lim_{OP}\, \frac{(P_{12}\, \dots\, P_{51})}{\textrm{det}^{\prime}(P_{ij})} \, .
\end{align}

Using  appendix \ref{slcrs} and eqn. (\ref{xi5forsoftlimit}) we can deduce that $\xi_{5}\, \rightarrow\, \xi_{5}^{soft}$, where,  
\begin{align}
\xi_{5}^{soft}\, =\, \lim_{\vert k_{5} \vert\rightarrow\, 0}\, \xi_{5} \, .
\end{align}
Finally we can combine equations (\ref{i5inversesoft}, \ref{i5soft}) to get the soft recursion relation for double scaled SYM correlators,  ${\cal A}^{\textrm{flat}}_{5}$ and ${\cal A}^{\textrm{flat}}_{4}$:
\begin{align}\label{stinsym}
&{\cal A}^{\textrm{flat}}_{5} \sqrt{\textrm{det}^{\prime}(P)}(\, \xi^{soft}_{5}, \vec{\sigma}_{\textrm{soft}}\, ) \nonumber \\& \qquad =\, \int \eta^{15}\, K_{0}(2\xi_{5}^{soft}\, \eta)\, \left(2 - \sum_{i=1}^{4} k_{i}\, \partial k_{i} \right)\, \hat{I}_{4}^{-1}\left(\, \sqrt{\textrm{det}^{\prime}(P_{4})}\, {\cal A}^{\textrm{flat}}_{4}\, \right)
\end{align}
We remind the reader that
\begin{align}
{\cal A}_{4}^{\textrm{flat}}\, =\, {\cal A}_{4}^{\textrm{flat}}(\xi_{4}, \sigma_{2})
\end{align}
Eqn(\ref{stinsym}) is then a recursion relation for a 5 point $\frac{1}{2}$ BPS-correlator in the double scaling limit of planar ${\cal N} = 4$ SYM.

The result is simply a consequence of flat space holography for perturbative string amplitudes and soft factorization theorem which this amplitude satisfies. However, purely from the gauge theory perspective, the result is rather striking since it shows a relationship between 5 and 4 point amplitude when the cross ratios satisfy a non-linear constraint.

The recursion relation obtained above should be differentiated from $SL(2,Z)$ covariant recursion relations obtained in \cite{Green:2020eyj}. In this work, the authors derive a relation between $n$ and $n-1$ point correlators in the maximal $U(1)_{Y}$ violating (MUV) sector. They consider $n$ point correlators with $4$ stress tensor insertions and $n-4$ insertions of a Chiral Lagrangian operator ${\cal O}_{\tau}$ where $\tau$ is the complexified coupling of SYM theory. These recursion relations are proved to be dual to the bulk soft axio-dilaton theorem in the super-gravity approximation. On the other hand, equation (\ref{stinsym}) is in the bosonic sector and is dual to tree-level dilatonic string amplitude where $\alpha^{\prime}$ corrections are not suppressed.  In any case, a more detailed comparison between the two relations is desirable but is outside the scope of this work.

Finally, we note that the recursion relation obtained above is then a factorization theorem for Carrollian correlators. Since it is equivalent to the sub-leading soft dilaton theorem, we do expect this theorem to be associated to an infinity of conserved currents related to the Weyl component of the conformal BMS group, \cite{Haco:2017ekf}. However, a detailed exploration of such an ``IR-edge'' will be pursued elsewhere. 

\section{Double Scaling limit of AdS $\phi^{3}$ S-matrix as Carrollian Correlators}\label{opphicubed}

\subsection{The OP-like limit}
As alluded to in section (\ref{revofderivop4}), the double scaling limit can also be used to obtain conformal correlators on $\partial\, \textrm{AdS}_{d+1}$ which are dual to flat space amplitudes of local effective field theories. As an illustration, we consider a simple example of tree-level 4 point amplitude in AdS$_{4}$. In this case, the double scaling limit is simply $L\, \rightarrow\, \infty,\, \rho\, L\, =\, \textrm{fixed}$. Then, by simply substituting the 4 point flat space amplitude in eqn.(\ref{fin_npt}), we find, 
\begin{align}\label{phi3op4pt}
{\cal F}_{\phi^{3}}^{tree}(\tilde{\xi}, \sigma)\, \sim\, \frac{\lambda^{2}}{\tilde{\xi}_4^{7}}\, \big(\, 1 - \frac{1}{\sigma} - \frac{1}{1-\sigma}\, \big)
\end{align}
In the above, $\tilde{\xi}_4\, =\, \frac{L}{L_{0}}\rho$ \footnote{The tilde on $\xi$ is basically used to distinguish it from \eqref{eq:xi4_main} which has angular dependence through $\sigma$.} and $\sigma$ are held fixed as $L\, \rightarrow\, \infty$. $L_{0}$ is a fixed length scale and can be chosen as $\Lambda_{UV}^{-1}$, where $\Lambda_{UV}$ is the UV cut off scale for the effective field theory.

\smallskip

This double scaling limit is inspired by the OP limit of the bulk amplitude. It is important to however notice that since the AdS/CFT duality implies that $\frac{L}{l_{s}}\, \sim\, N^{\frac{1}{4}}$, the OP limit of the string amplitude in AdS$_{5}\, \times\, S^{5}$ involves scaling of $\rho$ entirely in terms of parameters of the dual gauge theory. However, for non-holographic dual CFTs whose correlators can be written in terms of bulk correlators of local QFTs such as e.g. $\phi^{3}$ theory, there is no intrinsic string scale in which we measure units of energy and hence $\rho\, \rightarrow\, 0$ has no intrinsic formulation in terms of parameters of the boundary theory. 
We can now use eqn (\ref{xi4sq}) to re-write the RHS of eqn(\ref{phi3op4pt}) as a 4 point Carrollian correlator on ${\I}^{+}$.
\begin{align}
\langle\, \tilde{O}(u_{1}, \Omega_{1})\, \dots\, \tilde{O}(u_{4}, \Omega_{4})\, \rangle\,  \sim\, \frac{\lambda^{2}}{\left(\tilde{\xi}_4^0\right)^{7}}\, \bigg(\, 1 - \frac{1}{\sigma} - \frac{1}{1-\sigma}\, \bigg) \, ,
\end{align}
where $\tilde{\xi}_4^0$ is proportional to the $u$ dependent numerator of equation \eqref{xi40}.
Several comments are in order at this stage.
\begin{itemize}
\item The integral kernel that leads to (\ref{phi3op4pt}) arises from the fact that the conformal dimension of the massless scalar field in $\textrm{AdS}_{d+1}$ is $\triangle = d$. However, this dimension has no intrinsic meaning in the $L\, \rightarrow\, \infty$ limit. This immediately implies that the flat space amplitude obtained in the OP limit can not be interpreted as a boundary correlator of the radiative scalar field in Mink$_{4}$. The operator $\tilde{O}(u_{1}, \Omega_{1})$ is not of the form
\begin{align}
\tilde{O}(u_{1}, \Omega_{1})\, =\, \int\, dE\, e^{-i E u}\, \tilde{\phi}(E,\Omega_{1}) \, ,
\end{align}
where $\tilde{\phi}$ is the Fourier transform of the bulk scalar field. The field with engineering dimension $\triangle$ in flat space is, 
\begin{align}
\tilde{O}_{\triangle}(u_{1}, \Omega_{1})\, =\, \int\, dE\, E^{\triangle - 1}\, e^{-i E u}\,  \tilde{\phi}(E, \Omega_{1}) \, .
\end{align}
For $d=3$ we thus see that
\begin{align}
\tilde{O}_{3}(u_{1}, \Omega_{1})\, =\, -\partial_{u_{1}}^{2}\, \phi_{\textrm{rad}}(u_{1}, \Omega_{1}).
\end{align}
Hence, in the case of Mink$_{4}$, the double scaling limit can be understood as \emph{defining} Carrollian correlator $\langle \partial_{u_{1}}^{2}\, \phi(u_{1}, \Omega_{1})\, \dots\, \partial_{u_{4}}^{2}\, \phi(u_{4}, \Omega_{4})\, \rangle$.

In general dimensions however, there exists no Carrollian field at ${\I}^{+}$ which corresponds to $\tilde{O}_{\triangle = d}(u,\Omega)$. This is because in $d+1$ dimensions, the radiative scalar field at ${\I}^{+}$ has the following expansion in terms of Fourier modes of the bulk field. 
\begin{align}
\phi_{\textrm{rad}}(u,\, \Omega)\, =\, \int_{0}^{\infty}\, dE\, e^{-i E u}\, E^{\frac{d-3}{2}}\, \tilde{\Phi}(E, \Omega).
\end{align}
Hence
\begin{align}
\tilde{O}_{\Delta=d}\, =\, (i\partial_{u})^{ (d-1) - \frac{d-3}{2} } \phi_{\textrm{rad}}(u, \Omega) = (i\partial_u)^{\frac{d+1}{2}} \phi_{\text{rad}}(u,\Omega) \, . 
\end{align}
Thus in general dimension, the Carrollian correlator obtained via OP limit can not be defined intrinsically in flat space-time using local Carrollian field operators at ${\I}^{+}\, \cup\, {\I}^{-}$.
\item In this case, the integral transform in eqn.(\ref{fin_npt}) has no intrinsic interpretation in flat space. This transform arises precisely in the double scaling limit. Hence the OP limit produces Carrollian field theory only via flat space limit of AdS amplitudes as opposed to having an intrinsic flat space derivation.
\item This simple example already reveals something intriguing. That is, we can define an entire class of Carrollian CFTs via the double scaling limit of amplitudes in a local quantum field theory in AdS$_{4}$. As can already be seen from the example analyzed in this paper, the correlation functions so obtained have rather unusual pole structure and analytic properties which, to the best of our understanding, have not been witnessed in known examples of Carrollian QFTs.
\end{itemize}

\subsection{Carroll correlators from Witten diagrams}
\begin{figure}[tbp]
	\centering
    \hspace{1.5cm}
	\includegraphics[width=0.5\textwidth]{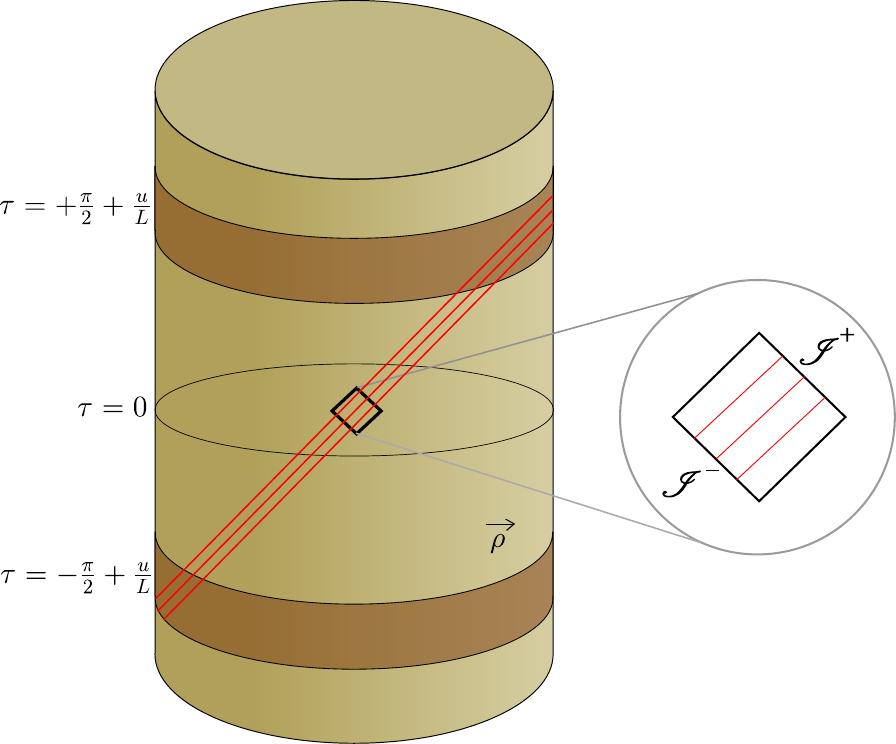}
	\caption{The Carrollian approach to flat limit of AdS.}
    \label{fig:flatads}
\end{figure}

The limit from AdS to flatspace and the rewriting of bulk Witten diagrams in terms of Carrollian correlators has been understood recently in \cite{Bagchi:2023fbj,Bagchi:2023cen}. This follows the logic laid out in the seminal paper by Penedones in \cite{Penedones:2010ue} where the operators on the boundary CFT were inserted on time-bands around $\tau_p = \pm \frac{\pi}{2} + \frac{u}{L}$. It was shown in \cite{Bagchi:2023fbj,Bagchi:2023cen} (following similar arguments for Celestial holography in \cite{deGioia:2022fcn,deGioia:2023cbd}) that focussing on the time bands on the dual CFT side naturally led to 3d Carrollian correlation functions when the flat limit was taken on AdS$_4$. The piece $\frac{u}{L}$ played a crucial role in zooming into the Minkowski diamond at the centre of AdS and projecting the dual CFT operators on $\I^\pm$ on this diamond as described in figure \ref{fig:flatads}.

In section~\ref{null_map}, we have already seen that the OP kinematics forces the dual operators to lie on the time bands at $\tau_p = \pm \frac{\pi}{2} + \frac{u}{L}$. Given the similarity of the picture of the double scaling limit advocated in this paper and this time-band point of view, we now make some concrete observations. Let us focus on scalar $\phi^3$ interactions in AdS$_4$ and following \cite{Bagchi:2023cen}, we can write the four-point exchange interaction of identical scalars $O$ with weight $\Delta$ as 
\begin{equation}\label{eq:4ptexchange_carroll}
    \begin{split}
        \langle O_{\Delta_1}(P_1) O_{\Delta_2}(P_2) O_{\Delta_3}(P_3) O_{\Delta_4}(P_4) \rangle = \mathcal{B} &\, \prod_{i=1}^4 (\sigma^*_i)^{\Delta_i-1} \frac{\delta(|z-\bar{z}|)}{z^2_{12} \bar{z}_{12} \bar{z}_{13} z_{24} \bar{z}_{24}} \prod_{i=1}^4 \indicator_{[0,1]}(\sigma^*_i) \\
        &\times \frac{1}{(\sigma^*_1 u_1 + \sigma^*_2 u_2 -\sigma^*_3 u_3 - \sigma^*_4 u_4)^{\Delta_1+\Delta_2+\Delta_3+\Delta_4-6}} \, ,
    \end{split}
\end{equation}
where $\mathcal{B}$ is a constant, $\indicator_{[0,1]}(x)$ is the indicator function defined as
\begin{equation}
    \indicator_{[0,1]}(x) = \begin{cases}
        1, \quad \text{if} ~ x \in [0,1] \\
        0, \quad \text{otherwise}
    \end{cases}
\end{equation}
$z, \bar{z}$ are the standard conformal ratios
\begin{equation}
    z = \dfrac{z_{12} z_{34}}{z_{13} z_{24}} \, , \qquad \bar{z} = \dfrac{\bar{z}_{12} \bar{z}_{34}}{\bar{z}_{13} \bar{z}_{24}} \, .
\end{equation}
$\sigma^*_i$ are given by
\begin{equation}\label{eq:sigmai}
    \begin{split}
        \sigma^*_1 = \frac{1}{D} \frac{z_{24} \bar{z}_{34}}{z_{12} \bar{z}_{13}} \, , \quad &\sigma^*_2 = - \frac{1}{D} \frac{z_{34} \bar{z}_{14}}{z_{23} \bar{z}_{12}} \, , \quad \sigma^*_3 = -\frac{1}{D} \frac{z_{24} \bar{z}_{14}}{z_{23} \bar{z}_{13}} \, , \quad \sigma^*_4 = \frac{1}{D} \, , \\
        &D = 2 \frac{z_{24} \bar{z}_{34}}{z_{12} \bar{z}_{13}}-2\frac{z_{34} \bar{z}_{14}}{z_{23} \bar{z}_{12}} \, .
    \end{split}
\end{equation}
We use OP kinematics to simplify our answer. We know: $\sum_{i=1}^4 \bf{e}_i = 0$. For the scattering configuration $12 \to 34$,  \eqref{mce} can be implemented as $\bf{e}_1 = -\bf{e}_2$ and $\bf{e}_3 = - \bf{e}_4$. The delta function $\delta(|z - \bar{z}|)$ imposes the constraint that all 4 insertion points lie on the great circle. These facts let us parametrize the points as follows:
\begin{equation}
    \begin{split}
        (z_1,\bar{z}_1) = \left(a, \frac{1}{a} \right) \implies (z_2,\bar{z}_2) = \left(-a, - \frac{1}{a} \right) \, , \quad |a|=1 \nonumber\\
        (z_3,\bar{z}_3) = \left(b, \frac{1}{b} \right) \implies (z_4,\bar{z}_4) = \left(-b, - \frac{1}{b} \right) \, , \quad |b|=1
    \end{split}
\end{equation}
Hence, we get: $\sigma^*_1 = \sigma^*_2 = \sigma^*_3 = \sigma^*_4 = \frac{1}{4}$. Substituting this into \eqref{eq:4ptexchange_carroll}, the dependence on $u_i$ in the denominator simplifies
\begin{equation}\label{eq:denomsimplify}
    (\sigma^*_1 u_1 + \sigma^*_2 u_2 -\sigma^*_3 u_3 - \sigma^*_4 u_4)^{4\Delta-6} = \left( \frac{1}{4} (u_1+u_2-u_3-u_4) \right)^{4\Delta-6} \, .
\end{equation}
We see that the pole structure is identical to \eqref{phi3op4pt}, if we take 
\begin{align}
    \Delta = \frac{13}{4}. 
\end{align}
The point of this exercise is the following. The correlators obtained by the OP method and the ones obtained as flat limits of Witten diagrams are both Carrollian conformal correlators and have to satisfy Carrollian conformal Ward identities. The pole structure in $u$ of these two objects would be completely fixed just by Carroll Ward identities. This would be enough to fix the weights of operators in the 4-point identical exchange. The OP limit thus seems to generate Carrollian CFT operators with strange fractional weight, indicating that these are possibly highly non-local operators from the point of view of a Carroll CFT. This is consistent with the recent observation that the dual Carrollian description may be non-local \cite{Cotler:2025npu}. It is likely that these operators and correlation functions define a Carroll CFT which would not obtainable by a conventional Carroll limit of a relativistic CFT where one can e.g. obtain Lagrangian descriptions of Carroll CFTs by a method of expansions about the vanishing speed of light limit. The OP limit thus has provided us a very new and very non-trivial way of generating a Carroll CFT from a relativistic one, far removed (it seems) from conventional vanishing lightspeed limits.  

%%%%%%%%%%%%%%%%%%%%%%%%%%%%%%%%%%%

\section{Conclusion}
\label{conclusion}
Most advances in flat space holography in the past decade have emerged in the so-called bottom-up approach that reconstructs bulk observables (such as the S-matrix) of gauge theories and perturbative gravity in flat space-time in terms of correlators on celestial $S^{2}$ or ${\I}^{+}$. The soft (or soft collinear) factorization theorems then lead to existence of an infinite dimensional current algebra which can then be used to write a class of boundary Celestial or Carrollian CFTs.

On the other hand, holography is a statement about quantum gravity and as such requires (at the very least) a precise conjecture that identifies the S-matrix (or inclusive observables) of quantum gravity in AFS with correlators in a non-gravitational theory that lives on a manifold of co-dimension $\geq\, 1$. In fact, several non-perturbative proposals including the BFSS matrix model \cite{Banks:1996vh}, the so called Matrix string theory and IKKT Matrix model \cite{Ishibashi:1996xs} have been proposed as candidates for flat space holography. Even though a complete dual description of string theory in AFS is likely to be far more nuanced (see \cite{Sen:2025oeq, Sen:2025ljz, Sen:2025bmj} for details), all of the models mentioned above offer tantalizing hints into how physics at the Planck scale may admit a dual description in terms of certain large $N$ matrix degrees of freedom. In this light, \emph{a} flat space limit of the AdS/CFT duality, proposed and analyzed in \cite{Okuda:2010ym} offers a complementary window into a specific corner of flat space holography. It is a corner that identifies tree-level string amplitudes at finite $\alpha^{\prime}$ with double scaling limit of correlation functions in ${\cal N} = 4$ SYM theory.

However, the doubly-scaled sector of SYM theory lives on $\partial$AdS$_{5}$ and hence this model of holography, based faithfully on AdS/CFT duality in its purest form, appears to be far removed from the recent advancement in bottom-up approaches. Our primary goal in this paper has been to take first steps towards showing that these two routes are not as distinct as they first appear.

Generalizing the result of \cite{Okuda:2010ym} to higher point tree-level dilatonic string amplitudes in ${\mathbb{R}}^{1,9}$, we have shown that the OP limit of SYM correlators can be used as basic data to define a set of  correlation functions in ${\I}^{+}$ that satisfy global Carrollian Ward identities. Specifically, at four point we proved that identifying double scaling limit of correlation functions (where external points converge to a common light cone on a Minkowski space-time) with correlators parametrized in terms of co-ordinates on ${\I}^{+}$ indeed defines observables of a strongly coupled Carrollian CFT which admits a ``genus'' expansion in terms of $g_{YM}$. We have already stressed that this Carrollian CFT is an effective theory that sits at the infinite coupling point of ${\cal{N}}=4$ SYM and it is unlikely that this is a usual Carroll CFT. Given this genus expansion, it is plausible that this is a topological theory. This is of course a preliminary remark which requires further detailed study. However, this simple observation of the genus expansion may have deeper implications for Carrollian holography.

 We then use the double scaling OP limit to derive new non-perturbative recursion relations in the dual gauge theory. In particular, we recast the tree-level dilaton sub-leading soft theorem in perturbative string theory as a recursion relation in the doubly scaled limit of ${\cal N} = 4$ SYM theory. We expect these recursion relations to correspond to an emergent infinite dimensional symmetry of the Carrollian CFT, plausibly corresponding to the Weyl subgroup of the Conformal BMS group.

 We then apply the lessons learned from the study of OP limit to bottom up approaches of flat space holography. In other words, starting with the flat space limit of the tree-level AdS$_{4}$ amplitude in the massless $\phi^{3}$ theory, we derive the corresponding Carrollian correlator on ${\I}^{+}$. Since the scaling dimension of the external state remains unchanged under the OP limit, as we show in section~\ref{opphicubed}, it seems unlikely that the resulting Carrollian CFT admits a Lagrangian description in terms of local radiative fields on ${\I}^{+}$. In this sense, the essential idea advocated in \cite{Okuda:2010ym} can be used to construct a potentially new class of Carrollian CFTs. It will be interesting, for example, to consider the large $N$ limit of the 4 point amplitudes in  $SU(N)$ QCD with massless adjoint matter so that $\frac{x}{t}\, \rightarrow\, N$ and to use it to define connected correlators of \emph{a} Carrollian CFT on ${\I}^{+}$.

Thus in a nutshell our work should be viewed as an attempt at the revival of the beautiful and under-explored corner for flat space holography that is reached via the OP limit. In this light, several immediate questions remain to be answered. The extension of the flat-space limit for AdS graviton amplitudes in AdS$_{5}\, \times\, S^{5}$ remains open. This is more than merely generalization for its own sake. Having holographic dual of tree-level graviton amplitude will then naturally lead to a discovery of new recursion formulae in ${\cal N} = 4$ SYM theory associated to super-translation and super-rotation Ward identities in the dual gauge theory. Such a tower of symmetries would be a striking vindication of the OP limit. The extension of the OP limit to perturbative string amplitudes in AdS$_{5}\, \times\, S^{5}$ at higher order in $g_{s}$ expansion remains completely open as the  KK modes in $S^{5}$ will propagate along the loops even if the external kinematics was supported only in ${\mathbb{R}}^{1,3}$. To ensure that the Carrollian correlators we have obtained exist as correlation functions of a theory which admits a genus expansion, proof of the existence of an OP limit at $O(g_{s})$ appears essential. 

Finally, there is some debate in the community whether the dual of the flat limit of AdS$_5 \times$S$^5$ should be a Carrollian theory that lives in 4 or 9 dimensions. The OP limit investigated in this paper points to a 4d dual Carroll CFT. However we note that, the kinematical data of the flat space S-matrix in this analysis lies in the 5 dimensional hyperplane of ${\mathbb{R}}^{1,9}$. Perhaps more importantly, the OP limit so far only exists for tree level amplitudes so that the 5 dimensional kinematics trivializes the KK modes on S$^5$ and exploring higher genus amplitudes would be vital to understanding if the putative dual can indeed be packaged as a 4d Carroll CFT or has an emergent 9d structure perhaps more in tune with expectations of a dual to a 10d string theory on AFS. 
\subsection*{Acknowledgements}
We thank Sujay Ashok,  Daniel Grumiller, Kristan Jensen, Arthur Lipstein, Shiraz Minwalla, Suvrat Raju, Ashoke Sen, Joan Simon, and Ronak Soni for illuminating discussions. 

AB thanks K. Narayan for an invitation to CMI and for hospitality in October 2025 during which the project was initially formulated. AB's research is partially supported by an ANRF MATRICS grant (ANRF/ARGM/2025/000653/MTR). AB also gratefully acknowledges the support of the Gireesh Jankinath Chair Professorship at IIT Kanpur. AL thanks Qubits and the spacetime unit at Okinawa Institute of Science and Technology for their warm hospitality, where part of this work was done. PD is supported by an NSERC discovery grant.

\newpage
\appendix
\makeatletter
\renewcommand{\@seccntformat}[1]{%
  \ifstrequal{#1}{section}{Appendix~\thesection:\quad}{\csname the#1\endcsname\quad}}
\makeatother
\section{Mapping $s^{0}_{ij}$ to $\vec{\sigma}$.}
\label{app_mandelstam}
In this section, we identify the map between the 4 independent Mandelstam angle variables and the 4 cross ratios which remain invariant under the OP limit. Let
\begin{align}
s^{0}_{ij}\, =\, \frac{{s}_{ij}}{s_{12}}\, \forall\, (i,j)\, \in\, \{\, (1,2),\, (2,3),\, (3,4),\, (4,5),\, (5,1)\, \}
\end{align}
The choice of $s_{12}$ as the ``energy variable'' is just for convenience, as the basis of angle variables is $\{s^{0}_{23}, s^{0}_{34}, s^{0}_{45}, s^{0}_{51}\}$. However, in principle, one could choose any linear combination of Mandelstam invariants as the ``energy variable''.\footnote{This split of Mandelstam invariants into one scale variable and the rest of dimensionless ratio of Mandelstam variables is known as angle-scale parameterization in the S-matrix literature, \cite{Brown:2011pj}.}

\medskip

The boundary insertions for the SYM correlator are in the projective space with coordinates $[P_{i}]\, :=\, \{\, P_{i}\, \sim\, \lambda P_{i}\, \}$. We start by choosing a specific ``gauge-fixed'' basis for these equivalence classes such that
\begin{align}\label{pii+1=1}
     \tilde{P}_{i,i+1} = -2 \tilde{P}_{i} \cdot \tilde{P}_{i+1}\, =\, 1
\end{align}
To see what this gauge fixing condition implies, let us start with any representative $P_{i}$ and define, 
\begin{align}\label{eq:softproj}
\tilde{P}_{i} := \gamma_{i}\, P_{i}
\end{align}
Then clearly, eqn.(\ref{pii+1=1}) implies that
\begin{align}
\gamma_{i}\gamma_{i+1}\, =\, \frac{1}{P_{i} \cdot P_{i+1}}\,\forall\, 1\leq\, i\, \leq\, 5 \nonumber\\
\end{align}
In turn, we can use this to express the matrix elements of $[P_{i} \cdot P_{j}]$ in terms of the 5 (dependent) cross ratios 
\begin{align}\label{s1to5P1to5}
\sigma_{1} = \frac{P_{13}P_{45}}{P_{15} P_{34}}\nonumber\\
\sigma_{2}\, =\, \frac{P_{14} P_{23}}{P_{12} P_{34}}\nonumber\\
\sigma_{3}\, =\, \frac{P_{24} P_{35}}{P_{23} P_{45}}\nonumber\\
\sigma_{4}\, =\, \frac{P_{25}P_{34}}{P_{23} P_{45}}\nonumber\\
\sigma_{5}\, =\, \frac{P_{35} P_{14}}{P_{15} P_{34}}
\end{align} 
We note that our choice of cross ratios is such that only $\sigma_{2}$ is independent of $P_{5}$.  
It can be immediately checked that the matrix elements of $[\tilde{P}_{ij}]$ are simple rational functions of these cross ratios: 
\begin{align}
\tilde{P}_{13} =\, \sigma_{1},\, \qquad \tilde{P}_{14}\, =\, \sigma_{2}\, \qquad \tilde{P}_{25} = \sigma_4, \\
\tilde{P}_{24} = \frac{\sigma_2 \sigma_3}{\sigma_5}, \qquad \tilde{P}_{35} = \frac{\sigma_5}{\sigma_2}, \qquad \textrm{etc.}
\end{align}
We note that if we simply take the strict limit where $\textrm{det}(P_{ij})\, =\, 0$ then in this limit
\begin{align}
\sigma_{5}\, =\, \sigma_{5}(\sigma_{1}, \dots, \sigma_{4}) = \sigma_{5}^{(0)}.
\end{align}
However, as we approach the OP limit, this relation is satisfied up to the leading order correction parameterized by the vanishing eigen-value $\lambda_{0}$.
\begin{align}
\sigma_{5}\, =\, \sigma_{5}^{(0)}\, +\, \delta\sigma_{5},
\end{align}
where $\delta \sigma_{5}\, \sim\, \frac{1}{\sqrt{N}}$. Thus, as the 5 points $x_{1}, \dots, x_{5}$ approach the light cone of a common point $x_{0}$, we have
\begin{align}
\textrm{det}(\tilde{P}_{ij})\, =\, \frac{\partial{\textrm{det}(\tilde{P}_{ij})}}{\partial \sigma_{5}}\vert_{\sigma_{5}^{(0)}}\, \delta\sigma_{5}\nonumber\\
\delta\sigma_{5}\, =\, \lambda_{0} \, \textrm{det}^{\prime}(\tilde{P}_{ij})\, \frac{1}{\frac{\partial\, \textrm{det}(\tilde{P}_{ij})}{\partial \sigma_{5}}\vert_{\sigma_{5}^{(0)}}\, }
\end{align}
Thus, 
\begin{align}
\xi^2_{5}\, \sim\, -\, N^{\frac{1}{2}}\, \lambda_{0}\, \sim\, -\, N^{\frac{1}{2}}\, \delta\sigma_{5}
\end{align}
This shows that $\xi_{5}$ is a measure of the rate at which $\sigma_{5}\, \rightarrow\, \sigma_{5}(\vec{\sigma})$.

\medskip

In order to finally arrive at the precise mapping between $s^{0}_{ij}$ and $\vec{\sigma}$, 
 we need to find a map between the projective representations of \eqref{eq:softproj} and the OP limit of \S\ref{npt_gen}. Let $(\mu_{1}, \dots, \mu_{5})^{T}$ be the null eigenvector of $\tilde{P}_{ij}$ in the OP limit. Without loss of generality, we can assume that $\mu_{1}\, \neq\, 0$. Let, 
\begin{align}
\mu_{1i}\, :=\, \frac{\mu_{i}}{\mu_{1}}\, \quad \forall\, 2\, \leq\, i\, \leq\, 5.
\end{align}
As all the matrix elements of the matrix $\tilde{P}_{ij}$ belong to the set $\vec{\sigma}$, we have the following.
\begin{align}
\mu_{1i} = \mu_{1i}(\vec{\sigma})
\end{align}
Okuda and Penedones relate the flat space momentum variables to a specific representative $P^{\prime}_{i}\, \in\, [P_{i}]$ such that in the $P^{\prime}_{ij}$ basis the null vector is $\frac{1}{\sqrt{n}}(1,\, \dots,\, 1)^{T}$. It is in this gauge-fixed representative of $[P_{i}]$ that the boundary co-ordinates are mapped to physical massless momenta via
\begin{align}\label{kietapprimei}
k_{i}\, =\, -\frac{2\eta}{l_{s}}\, P_{i}^{\prime}.
\end{align}
The relationship between $\{\tilde{P}_{i}\}$ and $P^{\prime}_{i}$ can be immediately written as
\begin{align}\label{tildepprimep}
P^{\prime}_{i} = \mu_{i}(\vec{\sigma})\, \tilde{P}_{i}
\end{align}
This is a mapping between two ``gauge fixed'' bases in the projective space  and therefore should not be misunderstood as a trivial map in the equivalence class $[\tilde{P}_{i}]$.\\
Finally, combining equations (\ref{kietapprimei}, \ref{tildepprimep}) and $s^{0}_{ij} = \frac{l_{s}^{2}}{\eta^{2}}\, s_{ij}$, we finally obtain an important result. 
\begin{align}\label{sii1vsig}
\boxed{s^{0}_{i,i+1}\, \sim\, \mu_{i}(\vec{\sigma})\, \mu_{i+1}(\vec{\sigma})}
\end{align}
We have thus obtained an explicit map between the angle variables $\{s^{0}_{ij}\}$ of flat space kinematics and the conformal cross ratios.\\
\subsection*{Soft limit in the cross-ratios space}\label{slcrs}
We can now take the soft limit in the space of cross ratios. It is far more convenient for this purpose to start with the inverse mapping from $s_{ij}$ to $\vec{\sigma}$ which can be deduced using eqn.(\ref{s1to5P1to5}). For example,  
\begin{align}\label{sig1s13s45}
\sigma_{1}\, =\, \frac{s_{13} s_{45}}{s_{15} s_{34}}.
\end{align}
We will now use equations~\eqref{sii1vsig} and~\eqref{sig1s13s45} to derive the soft constraint on $\vec{\sigma}$.
Using momentum conservation, we can derive a linear relationship between 
\begin{align}
s_{13} = s_{45} - s_{12} - s_{23},\nonumber\\
\implies\, s_{13}^{soft}\, =\, -s_{12} - s_{23}
\end{align}
Thus,
\begin{align}
\sigma_{1}\, \rightarrow\, \sigma_{1}^{\textrm{soft}}
\end{align}
where using $s^{0}_{12}\, =\, 1$ we find that 
\begin{align}\label{eq:soft_sigma}
\sigma_{1}^{\textrm{soft}}\, =\, -\, (\, 1 + s^{0}_{23}\, )\, \frac{s^{0}_{45}}{s^{0}_{15}},\nonumber\\
\sigma_{2}^{\textrm{soft}}\, =\, \sigma_{2}\, =\, (s^{0}_{23})^{2},\nonumber\\
\sigma_{3}^{\textrm{soft}}\, =\, -\, \frac{1 + s^{0}_{23}}{s^{0}_{23}}\, \frac{s^{0}_{35}}{s^{0}_{45}},\nonumber\\
\sigma^{\textrm{soft}}_{4}\, =\, \frac{1}{s^{0}_{23}}\, \frac{s^{0}_{25}}{s^{0}_{45}}.
\end{align}
Hence, the soft limit corresponds to a variety in the space of cross ratios. This variety is a solution to the following equation,
\begin{align}\label{softvar}
-\frac{1 + \sqrt{\sigma_{2}}}{\sigma_{1}^{\textrm{soft}}}\, +\, \sigma_{4}^{\textrm{soft}}\, \sqrt{\sigma_{2}}\, -\, \frac{\sigma_{3}^{\textrm{soft}}\, \sqrt{\sigma_{2}}}{1\, +\, \sqrt{\sigma_{2}}}\, +\, 1\, =\, 0.
\end{align}

\bibliography{References}

@article{Witten:1998qj,
    author = "Witten, Edward",
    title = "{Anti-de Sitter space and holography}",
    eprint = "hep-th/9802150",
    archivePrefix = "arXiv",
    reportNumber = "IASSNS-HEP-98-15",
    doi = "10.4310/ATMP.1998.v2.n2.a2",
    journal = "Adv. Theor. Math. Phys.",
    volume = "2",
    pages = "253--291",
    year = "1998"
}

@article{Dutta:2026etj,
    author = "Dutta, Sudipta",
    title = "{Massive fields in 3D Minkowski space and boundary correlators}",
    eprint = "2606.25803",
    archivePrefix = "arXiv",
    primaryClass = "hep-th",
    month = "6",
    year = "2026"
}

@article{Zheng:2026onv,
    author = "Zheng, Yu-fan",
    title = "{Carrollian Dictionary for Massive Particles at Null Infinity}",
    eprint = "2608.11945",
    archivePrefix = "arXiv",
    primaryClass = "hep-th",
    month = "8",
    year = "2026"
}

@article{Banks:1996vh,
    author = "Banks, Tom and Fischler, W. and Shenker, S. H. and Susskind, Leonard",
    title = "{M theory as a matrix model: A conjecture}",
    eprint = "hep-th/9610043",
    archivePrefix = "arXiv",
    reportNumber = "RU-96-95, SU-ITP-96-12, UTTG-13-96",
    doi = "10.1103/PhysRevD.55.5112",
    journal = "Phys. Rev. D",
    volume = "55",
    pages = "5112--5128",
    year = "1997"
}

@article{Ishibashi:1996xs,
    author = "Ishibashi, N. and Kawai, H. and Kitazawa, Y. and Tsuchiya, A.",
    title = "{A Large N reduced model as superstring}",
    eprint = "hep-th/9612115",
    archivePrefix = "arXiv",
    reportNumber = "KEK-TH-503",
    doi = "10.1016/S0550-3213(97)00290-3",
    journal = "Nucl. Phys. B",
    volume = "498",
    pages = "467--491",
    year = "1997"
}

@article{Melton:2026tdw,
    author = "Melton, Walker and Michaelsen, Kyrill and Ruzziconi, Romain",
    title = "{Observing Massive Scattering from Null Infinity}",
    eprint = "2606.27421",
    archivePrefix = "arXiv",
    primaryClass = "hep-th",
    month = "6",
    year = "2026"
}

@article{Barnich:2012rz,
	author = "Barnich, Glenn and Gomberoff, Andr\'es and Gonz\'alez, Hern\'an A.",
	title = "{Three-dimensional Bondi-Metzner-Sachs invariant two-dimensional field theories as the flat limit of Liouville theory}",
	eprint = "1210.0731",
	archivePrefix = "arXiv",
	primaryClass = "hep-th",
	doi = "10.1103/PhysRevD.87.124032",
	journal = "Phys. Rev. D",
	volume = "87",
	number = "12",
	pages = "124032",
	year = "2013"
}

@article{Cotler:2024xhb,
	author = "Cotler, Jordan and Jensen, Kristan and Prohazka, Stefan and Raz, Amir and Riegler, Max and Salzer, Jakob",
	title = "{Quantizing Carrollian field theories}",
	eprint = "2407.11971",
	archivePrefix = "arXiv",
	primaryClass = "hep-th",
	doi = "10.1007/JHEP10(2024)049",
	journal = "JHEP",
	volume = "10",
	pages = "049",
	year = "2024"
}

@article{Bagchi:2023fbj,
	author = "Bagchi, Arjun and Dhivakar, Prateksh and Dutta, Sudipta",
	title = "{AdS Witten diagrams to Carrollian correlators}",
	eprint = "2303.07388",
	archivePrefix = "arXiv",
	primaryClass = "hep-th",
	doi = "10.1007/JHEP04(2023)135",
	journal = "JHEP",
	volume = "04",
	pages = "135",
	year = "2023"
}

@article{Bagchi:2022emh,
	author = "Bagchi, Arjun and Banerjee, Shamik and Basu, Rudranil and Dutta, Sudipta",
	title = "{Scattering Amplitudes: Celestial and Carrollian}",
	eprint = "2202.08438",
	archivePrefix = "arXiv",
	primaryClass = "hep-th",
	doi = "10.1103/PhysRevLett.128.241601",
	journal = "Phys. Rev. Lett.",
	volume = "128",
	number = "24",
	pages = "241601",
	year = "2022"
}

@article{Donnay:2022aba,
	author = "Donnay, Laura and Fiorucci, Adrien and Herfray, Yannick and Ruzziconi, Romain",
	title = "{Carrollian Perspective on Celestial Holography}",
	eprint = "2202.04702",
	archivePrefix = "arXiv",
	primaryClass = "hep-th",
	doi = "10.1103/PhysRevLett.129.071602",
	journal = "Phys. Rev. Lett.",
	volume = "129",
	number = "7",
	pages = "071602",
	year = "2022"
}

@article{Donnay:2022wvx,
	author = "Donnay, Laura and Fiorucci, Adrien and Herfray, Yannick and Ruzziconi, Romain",
	title = "{Bridging Carrollian and celestial holography}",
	eprint = "2212.12553",
	archivePrefix = "arXiv",
	primaryClass = "hep-th",
	doi = "10.1103/PhysRevD.107.126027",
	journal = "Phys. Rev. D",
	volume = "107",
	number = "12",
	pages = "126027",
	year = "2023"
}

@article{Pasterski:2017kqt,
	author = "Pasterski, Sabrina and Shao, Shu-Heng",
	title = "{Conformal basis for flat space amplitudes}",
	eprint = "1705.01027",
	archivePrefix = "arXiv",
	primaryClass = "hep-th",
	doi = "10.1103/PhysRevD.96.065022",
	journal = "Phys. Rev. D",
	volume = "96",
	number = "6",
	pages = "065022",
	year = "2017"
}

@article{Bagchi:2023cen,
	author = "Bagchi, Arjun and Dhivakar, Prateksh and Dutta, Sudipta",
	title = "{Holography in flat spacetimes: the case for Carroll}",
	eprint = "2311.11246",
	archivePrefix = "arXiv",
	primaryClass = "hep-th",
	doi = "10.1007/JHEP08(2024)144",
	journal = "JHEP",
	volume = "08",
	pages = "144",
	year = "2024"
}

@article{Banerjee:2018gce,
	author = "Banerjee, Shamik",
	title = "{Null Infinity and Unitary Representation of The Poincare Group}",
	eprint = "1801.10171",
	archivePrefix = "arXiv",
	primaryClass = "hep-th",
	doi = "10.1007/JHEP01(2019)205",
	journal = "JHEP",
	volume = "01",
	pages = "205",
	year = "2019"
}

@article{Gary:2009ae,
	author = "Gary, Mirah and Giddings, Steven B. and Penedones, Joao",
	title = "{Local bulk S-matrix elements and CFT singularities}",
	eprint = "0903.4437",
	archivePrefix = "arXiv",
	primaryClass = "hep-th",
	reportNumber = "CERN-PH-TH-2009-035, NSF-KITP-09-35",
	doi = "10.1103/PhysRevD.80.085005",
	journal = "Phys. Rev. D",
	volume = "80",
	pages = "085005",
	year = "2009"
}

@article{Penedones:2010ue,
	author = "Penedones, Joao",
	title = "{Writing CFT correlation functions as AdS scattering amplitudes}",
	eprint = "1011.1485",
	archivePrefix = "arXiv",
	primaryClass = "hep-th",
	doi = "10.1007/JHEP03(2011)025",
	journal = "JHEP",
	volume = "03",
	pages = "025",
	year = "2011"
}

@article{Bagchi:2012xr,
	author = "Bagchi, Arjun and Detournay, St\'ephane and Fareghbal, Reza and Sim\'on, Joan",
	title = "{Holography of 3D Flat Cosmological Horizons}",
	eprint = "1208.4372",
	archivePrefix = "arXiv",
	primaryClass = "hep-th",
	reportNumber = "EMPG-12-18",
	doi = "10.1103/PhysRevLett.110.141302",
	journal = "Phys. Rev. Lett.",
	volume = "110",
	number = "14",
	pages = "141302",
	year = "2013"
}

@article{Donnay:2020guq,
	author = "Donnay, Laura and Pasterski, Sabrina and Puhm, Andrea",
	title = "{Asymptotic Symmetries and Celestial CFT}",
	eprint = "2005.08990",
	archivePrefix = "arXiv",
	primaryClass = "hep-th",
	reportNumber = "CPHT-RR022.042020",
	doi = "10.1007/JHEP09(2020)176",
	journal = "JHEP",
	volume = "09",
	pages = "176",
	year = "2020"
}

@article{Okuda:2010ym,
    author = "Okuda, Takuya and Penedones, Joao",
    title = "{String scattering in flat space and a scaling limit of Yang-Mills correlators}",
    eprint = "1002.2641",
    archivePrefix = "arXiv",
    primaryClass = "hep-th",
    doi = "10.1103/PhysRevD.83.086001",
    journal = "Phys. Rev. D",
    volume = "83",
    pages = "086001",
    year = "2011"
}

@article{Laddha:2022nmj,
    author = "Laddha, Alok and Prabhu, Siddharth G. and Raju, Suvrat and Shrivastava, Pushkal",
    title = "{Squinting at massive fields from infinity}",
    eprint = "2207.06406",
    archivePrefix = "arXiv",
    primaryClass = "hep-th",
    month = "7",
    year = "2022"
}

@article{H:2024cfo,
    author = "H, Anupam A. and Athira, V., P. and Paul, Priyadarshi and Raju, Suvrat",
    title = "{Interacting fields at spatial infinity}",
    eprint = "2405.20326",
    archivePrefix = "arXiv",
    primaryClass = "hep-th",
    doi = "10.1103/hm1z-qbfw",
    journal = "Phys. Rev. D",
    volume = "114",
    number = "2",
    pages = "026023",
    year = "2026"
}

@article{Have:2024dff,
    author = "Have, Emil and Nguyen, Kevin and Prohazka, Stefan and Salzer, Jakob",
    title = "{Massive carrollian fields at timelike infinity}",
    eprint = "2402.05190",
    archivePrefix = "arXiv",
    primaryClass = "hep-th",
    reportNumber = "UWThPh 2024-6",
    doi = "10.1007/JHEP07(2024)054",
    journal = "JHEP",
    volume = "07",
    pages = "054",
    year = "2024"
}

@article{Borthwick:2024skd,
    author = {Borthwick, Jack and Chantreau, Ma{\"e}l and Herfray, Yannick},
    title = "{Ti and Spi, Carrollian extended boundaries at timelike and spatial infinity}",
    eprint = "2412.15996",
    archivePrefix = "arXiv",
    primaryClass = "gr-qc",
    doi = "10.1088/1361-6382/ae0d3f",
    journal = "Class. Quant. Grav.",
    volume = "42",
    number = "20",
    pages = "205012",
    year = "2025"
}

@article{Liu:2025oom,
    author = "Liu, Wen-Bin and Long, Jiang",
    title = "{Extrapolating the massive fields to future timelike infinity}",
    eprint = "2508.15619",
    archivePrefix = "arXiv",
    primaryClass = "hep-th",
    doi = "10.1007/JHEP12(2025)042",
    journal = "JHEP",
    volume = "12",
    pages = "042",
    year = "2025"
}

@article{Polchinski:1999ry,
	Archiveprefix = {arXiv},
	Author = {Polchinski, Joseph},
	Eprint = {hep-th/9901076},
	Month = {1},
	Reportnumber = {NST-ITP-99-02},
	Title = {{S matrices from AdS space-time}},
	Year = {1999}}

@article{Susskind:1998vk,
	Archiveprefix = {arXiv},
	Author = {Susskind, Leonard},
	Doi = {10.1063/1.1301570},
	Editor = {Burgess, C. P. and Myers, Robert C.},
	Eprint = {hep-th/9901079},
	Journal = {AIP Conf. Proc.},
	Number = {1},
	Pages = {98--112},
	Title = {{Holography in the flat space limit}},
	Volume = {493},
	Year = {1999}}

@article{Maldacena:1997re,
	Archiveprefix = {arXiv},
	Author = {Maldacena, Juan Martin},
	Doi = {10.1023/A:1026654312961},
	Eprint = {hep-th/9711200},
	Journal = {Adv. Theor. Math. Phys.},
	Pages = {231--252},
	Reportnumber = {HUTP-97-A097, HUTP-98-A097},
	Title = {{The Large N limit of superconformal field theories and supergravity}},
	Volume = {2},
	Year = {1998}}

@article{Giddings:1999qu,
	Archiveprefix = {arXiv},
	Author = {Giddings, Steven B.},
	Doi = {10.1103/PhysRevLett.83.2707},
	Eprint = {hep-th/9903048},
	Journal = {Phys. Rev. Lett.},
	Pages = {2707--2710},
	Title = {{The Boundary S matrix and the AdS to CFT dictionary}},
	Volume = {83},
	Year = {1999}}

@article{Giddings:1999jq,
	Archiveprefix = {arXiv},
	Author = {Giddings, Steven B.},
	Doi = {10.1103/PhysRevD.61.106008},
	Eprint = {hep-th/9907129},
	Journal = {Phys. Rev. D},
	Pages = {106008},
	Title = {{Flat space scattering and bulk locality in the AdS / CFT correspondence}},
	Volume = {61},
	Year = {2000}}

@article{Fitzpatrick:2011hu,
	Archiveprefix = {arXiv},
	Author = {Fitzpatrick, A. Liam and Kaplan, Jared},
	Doi = {10.1007/JHEP10(2012)127},
	Eprint = {1111.6972},
	Journal = {JHEP},
	Pages = {127},
	Primaryclass = {hep-th},
	Reportnumber = {SLAC-PUB-14841},
	Title = {{Analyticity and the Holographic S-Matrix}},
	Volume = {10},
	Year = {2012}}

@article{Fitzpatrick:2011jn,
	Archiveprefix = {arXiv},
	Author = {Fitzpatrick, A. Liam and Kaplan, Jared},
	Eprint = {1104.2597},
	Month = {4},
	Primaryclass = {hep-th},
	Reportnumber = {SLAC-PUB-14507},
	Title = {{Scattering States in AdS/CFT}},
	Year = {2011}}

@article{Fitzpatrick:2011dm,
	Archiveprefix = {arXiv},
	Author = {Fitzpatrick, A. Liam and Kaplan, Jared},
	Doi = {10.1007/JHEP10(2012)032},
	Eprint = {1112.4845},
	Journal = {JHEP},
	Pages = {032},
	Primaryclass = {hep-th},
	Reportnumber = {SLAC-PUB-14979},
	Title = {{Unitarity and the Holographic S-Matrix}},
	Volume = {10},
	Year = {2012}}

@article{Raju:2012zr,
	Archiveprefix = {arXiv},
	Author = {Raju, Suvrat},
	Doi = {10.1103/PhysRevD.85.126009},
	Eprint = {1201.6449},
	Journal = {Phys. Rev. D},
	Pages = {126009},
	Primaryclass = {hep-th},
	Reportnumber = {HRI-ST-1201},
	Title = {{New Recursion Relations and a Flat Space Limit for AdS/CFT Correlators}},
	Volume = {85},
	Year = {2012}}

@article{Alday:2024yyj,
    author = "Alday, Luis F. and Nocchi, Maria and Ruzziconi, Romain and Yelleshpur Srikant, Akshay",
    title = "{Carrollian amplitudes from holographic correlators}",
    eprint = "2406.19343",
    archivePrefix = "arXiv",
    primaryClass = "hep-th",
    doi = "10.1007/JHEP03(2025)158",
    journal = "JHEP",
    volume = "03",
    pages = "158",
    year = "2025"
}

@article{Lipstein:2025jfj,
    author = "Lipstein, Arthur and Ruzziconi, Romain and Yelleshpur Srikant, Akshay",
    title = "{Towards a flat space Carrollian hologram from AdS$_{4}$/CFT$_{3}$}",
    eprint = "2504.10291",
    archivePrefix = "arXiv",
    primaryClass = "hep-th",
    doi = "10.1007/JHEP06(2025)073",
    journal = "JHEP",
    volume = "06",
    pages = "073",
    year = "2025"
}

@article{Paulos:2016fap,
	Archiveprefix = {arXiv},
	Author = {Paulos, Miguel F. and Penedones, Joao and Toledo, Jonathan and van Rees, Balt C. and Vieira, Pedro},
	Doi = {10.1007/JHEP11(2017)133},
	Eprint = {1607.06109},
	Journal = {JHEP},
	Pages = {133},
	Primaryclass = {hep-th},
	Reportnumber = {CERN-TH-2016-162},
	Title = {{The S-matrix bootstrap. Part I: QFT in AdS}},
	Volume = {11},
	Year = {2017}}

@article{Paulos:2016but,
    author = "Paulos, Miguel F. and Penedones, Joao and Toledo, Jonathan and van Rees, Balt C. and Vieira, Pedro",
    title = "{The S-matrix bootstrap II: two dimensional amplitudes}",
    eprint = "1607.06110",
    archivePrefix = "arXiv",
    primaryClass = "hep-th",
    reportNumber = "CERN-TH-2016-163",
    doi = "10.1007/JHEP11(2017)143",
    journal = "JHEP",
    volume = "11",
    pages = "143",
    year = "2017"
}

@article{Paulos:2017fhb,
    author = "Paulos, Miguel F. and Penedones, Joao and Toledo, Jonathan and van Rees, Balt C. and Vieira, Pedro",
    title = "{The S-matrix bootstrap. Part III: higher dimensional amplitudes}",
    eprint = "1708.06765",
    archivePrefix = "arXiv",
    primaryClass = "hep-th",
    reportNumber = "CERN-TH-2017-162",
    doi = "10.1007/JHEP12(2019)040",
    journal = "JHEP",
    volume = "12",
    pages = "040",
    year = "2019"
}

@article{He:2014laa,
    author = "He, Temple and Lysov, Vyacheslav and Mitra, Prahar and Strominger, Andrew",
    title = "{BMS supertranslations and Weinberg{\textquoteright}s soft graviton theorem}",
    eprint = "1401.7026",
    archivePrefix = "arXiv",
    primaryClass = "hep-th",
    doi = "10.1007/JHEP05(2015)151",
    journal = "JHEP",
    volume = "05",
    pages = "151",
    year = "2015"
}

@article{Lysov:2014csa,
    author = "Lysov, Vyacheslav and Pasterski, Sabrina and Strominger, Andrew",
    title = "{Low{\textquoteright}s Subleading Soft Theorem as a Symmetry of QED}",
    eprint = "1407.3814",
    archivePrefix = "arXiv",
    primaryClass = "hep-th",
    doi = "10.1103/PhysRevLett.113.111601",
    journal = "Phys. Rev. Lett.",
    volume = "113",
    number = "11",
    pages = "111601",
    year = "2014"
}

@article{He:2014cra,
    author = "He, Temple and Mitra, Prahar and Porfyriadis, Achilleas P. and Strominger, Andrew",
    title = "{New Symmetries of Massless QED}",
    eprint = "1407.3789",
    archivePrefix = "arXiv",
    primaryClass = "hep-th",
    doi = "10.1007/JHEP10(2014)112",
    journal = "JHEP",
    volume = "10",
    pages = "112",
    year = "2014"
}

@article{Campiglia:2014yka,
    author = "Campiglia, Miguel and Laddha, Alok",
    title = "{Asymptotic symmetries and subleading soft graviton theorem}",
    eprint = "1408.2228",
    archivePrefix = "arXiv",
    primaryClass = "hep-th",
    doi = "10.1103/PhysRevD.90.124028",
    journal = "Phys. Rev. D",
    volume = "90",
    number = "12",
    pages = "124028",
    year = "2014"
}

@article{Avery:2015gxa,
    author = "Avery, Steven G. and Schwab, Burkhard U. W.",
    title = "{Burg-Metzner-Sachs symmetry, string theory, and soft theorems}",
    eprint = "1506.05789",
    archivePrefix = "arXiv",
    primaryClass = "hep-th",
    reportNumber = "BROWN-HET-1674",
    doi = "10.1103/PhysRevD.93.026003",
    journal = "Phys. Rev. D",
    volume = "93",
    pages = "026003",
    year = "2016"
}

@article{Campiglia:2016hvg,
    author = "Campiglia, Miguel and Laddha, Alok",
    title = "{Subleading soft photons and large gauge transformations}",
    eprint = "1605.09677",
    archivePrefix = "arXiv",
    primaryClass = "hep-th",
    doi = "10.1007/JHEP11(2016)012",
    journal = "JHEP",
    volume = "11",
    pages = "012",
    year = "2016"
}

@article{Laddha:2017vfh,
    author = "Laddha, Alok and Mitra, Prahar",
    title = "{Asymptotic Symmetries and Subleading Soft Photon Theorem in Effective Field Theories}",
    eprint = "1709.03850",
    archivePrefix = "arXiv",
    primaryClass = "hep-th",
    doi = "10.1007/JHEP05(2018)132",
    journal = "JHEP",
    volume = "05",
    pages = "132",
    year = "2018"
}

@article{Strominger:2017zoo,
	Archiveprefix = {arXiv},
	Author = {Strominger, Andrew},
	Eprint = {1703.05448},
	Month = {3},
	Primaryclass = {hep-th},
	Title = {{Lectures on the Infrared Structure of Gravity and Gauge Theory}},
	Year = {2017}}

@article{Banerjee:2020zlg,
    author = "Banerjee, Shamik and Ghosh, Sudip and Paul, Partha",
    title = "{MHV graviton scattering amplitudes and current algebra on the celestial sphere}",
    eprint = "2008.04330",
    archivePrefix = "arXiv",
    primaryClass = "hep-th",
    doi = "10.1007/JHEP02(2021)176",
    journal = "JHEP",
    volume = "02",
    pages = "176",
    year = "2021"
}

@article{Guevara:2021abz,
    author = "Guevara, Alfredo and Himwich, Elizabeth and Pate, Monica and Strominger, Andrew",
    title = "{Holographic symmetry algebras for gauge theory and gravity}",
    eprint = "2103.03961",
    archivePrefix = "arXiv",
    primaryClass = "hep-th",
    doi = "10.1007/JHEP11(2021)152",
    journal = "JHEP",
    volume = "11",
    pages = "152",
    year = "2021"
}

@article{Strominger:2021mtt,
    author = "Strominger, Andrew",
    title = "{$w_{1+\infty}$ Algebra and the Celestial Sphere: Infinite Towers of Soft Graviton, Photon, and Gluon Symmetries}",
    eprint = "2105.14346",
    archivePrefix = "arXiv",
    primaryClass = "hep-th",
    doi = "10.1103/PhysRevLett.127.221601",
    journal = "Phys. Rev. Lett.",
    volume = "127",
    number = "22",
    pages = "221601",
    year = "2021"
}

@article{Himwich:2021dau,
    author = "Himwich, Elizabeth and Pate, Monica and Singh, Kyle",
    title = "{Celestial operator product expansions and w$_{1+\infty}$ symmetry for all spins}",
    eprint = "2108.07763",
    archivePrefix = "arXiv",
    primaryClass = "hep-th",
    doi = "10.1007/JHEP01(2022)080",
    journal = "JHEP",
    volume = "01",
    pages = "080",
    year = "2022"
}

@article{Banerjee:2021dlm,
    author = "Banerjee, Shamik and Ghosh, Sudip and Paul, Partha",
    title = "{(Chiral) Virasoro invariance of the tree-level MHV graviton scattering amplitudes}",
    eprint = "2108.04262",
    archivePrefix = "arXiv",
    primaryClass = "hep-th",
    doi = "10.1007/JHEP09(2022)236",
    journal = "JHEP",
    volume = "09",
    pages = "236",
    year = "2022"
}

@article{Raclariu:2021zjz,
	Archiveprefix = {arXiv},
	Author = {Raclariu, Ana-Maria},
	Eprint = {2107.02075},
	Month = {7},
	Primaryclass = {hep-th},
	Title = {{Lectures on Celestial Holography}},
	Year = {2021}}

@article{Pasterski:2021rjz,
	Archiveprefix = {arXiv},
	Author = {Pasterski, Sabrina},
	Doi = {10.1140/epjc/s10052-021-09846-7},
	Eprint = {2108.04801},
	Journal = {Eur. Phys. J. C},
	Number = {12},
	Pages = {1062},
	Primaryclass = {hep-th},
	Title = {{Lectures on celestial amplitudes}},
	Volume = {81},
	Year = {2021}}

@inproceedings{Pasterski:2021raf,
    author = "Pasterski, Sabrina and Pate, Monica and Raclariu, Ana-Maria",
    title = "{Celestial Holography}",
    booktitle = "{Snowmass 2021}",
    eprint = "2111.11392",
    archivePrefix = "arXiv",
    primaryClass = "hep-th",
    month = "11",
    year = "2021"
}

@article{Adamo:2021lrv,
    author = "Adamo, Tim and Mason, Lionel and Sharma, Atul",
    title = "{Celestial $w_{1+\infty}$ Symmetries from Twistor Space}",
    eprint = "2110.06066",
    archivePrefix = "arXiv",
    primaryClass = "hep-th",
    doi = "10.3842/SIGMA.2022.016",
    journal = "SIGMA",
    volume = "18",
    pages = "016",
    year = "2022"
}

@article{Costello:2022wso,
    author = "Costello, Kevin and Paquette, Natalie M.",
    title = "{Celestial holography meets twisted holography: 4d amplitudes from chiral correlators}",
    eprint = "2201.02595",
    archivePrefix = "arXiv",
    primaryClass = "hep-th",
    doi = "10.1007/JHEP10(2022)193",
    journal = "JHEP",
    volume = "10",
    pages = "193",
    year = "2022"
}

@article{Donnay:2023mrd,
    author = "Donnay, Laura",
    title = "{Celestial holography: An asymptotic symmetry perspective}",
    eprint = "2310.12922",
    archivePrefix = "arXiv",
    primaryClass = "hep-th",
    doi = "10.1016/j.physrep.2024.04.003",
    journal = "Phys. Rept.",
    volume = "1073",
    pages = "1--41",
    year = "2024"
}

@article{Melton:2024akx,
    author = "Melton, Walker and Sharma, Atul and Strominger, Andrew and Wang, Tianli",
    title = "{Celestial Dual for Maximal Helicity Violating Amplitudes}",
    eprint = "2403.18896",
    archivePrefix = "arXiv",
    primaryClass = "hep-th",
    doi = "10.1103/PhysRevLett.133.091603",
    journal = "Phys. Rev. Lett.",
    volume = "133",
    number = "9",
    pages = "091603",
    year = "2024"
}

@article{Mol:2024etg,
    author = "Mol, Igor",
    title = "{A Holographic Construction of MHV Graviton Amplitudes in Celestial CFT}",
    eprint = "2408.10944",
    archivePrefix = "arXiv",
    primaryClass = "hep-th",
    month = "8",
    year = "2024"
}

@article{Ghorai:2026qaj,
    author = "Ghorai, Nirmal and Paul, Partha and Suryanarayana, Nemani V.",
    title = "{Celestial dual of conformal gravity MHV amplitudes: an OPE analysis}",
    eprint = "2605.05363",
    archivePrefix = "arXiv",
    primaryClass = "hep-th",
    month = "5",
    year = "2026"
}

@article{Ghorai:2025ebc,
    author = "Ghorai, Nirmal and Paul, Partha and Suryanarayana, Nemani V.",
    title = "{Towards celestial CFT dual of 4d conformal gravity}",
    eprint = "2511.03669",
    archivePrefix = "arXiv",
    primaryClass = "hep-th",
    month = "11",
    year = "2025"
}

@article{Zhu:2026ofh,
    author = "Zhu, Bin",
    title = "{Topics in Celestial holography: A bottom-up perspective}",
    eprint = "2606.24285",
    archivePrefix = "arXiv",
    primaryClass = "hep-th",
    doi = "10.1016/j.physrep.2026.08.001",
    journal = "Phys. Rept.",
    volume = "1200",
    pages = "1--46",
    year = "2026"
}

@article{Mitra:2024ugt,
    author = "Mitra, Prahar",
    title = "{Celestial Conformal Primaries in Effective Field Theories}",
    eprint = "2402.09256",
    archivePrefix = "arXiv",
    primaryClass = "hep-th",
    month = "2",
    year = "2024"
}

@article{Bagchi:2025vri,
    author = "Bagchi, Arjun and Banerjee, Aritra and Dhivakar, Prateksh and Mondal, Saikat and Shukla, Ashish",
    title = "{The Carrollian kaleidoscope}",
    eprint = "2506.16164",
    archivePrefix = "arXiv",
    primaryClass = "hep-th",
    doi = "10.1140/epjc/s10052-026-15437-1",
    journal = "Eur. Phys. J. C",
    volume = "86",
    number = "4",
    pages = "429",
    year = "2026"
}

@article{Nguyen:2025zhg,
    author = "Nguyen, Kevin",
    title = "{Lectures on Carrollian holography}",
    eprint = "2511.10162",
    archivePrefix = "arXiv",
    primaryClass = "hep-th",
    doi = "10.1140/epjc/s10052-026-15952-1",
    journal = "Eur. Phys. J. C",
    volume = "86",
    number = "6",
    pages = "733",
    year = "2026"
}

@article{Ruzziconi:2026bix,
    author = "Ruzziconi, Romain",
    title = "{Carrollian physics and holography}",
    eprint = "2602.02644",
    archivePrefix = "arXiv",
    primaryClass = "hep-th",
    doi = "10.1016/j.physrep.2026.03.005",
    journal = "Phys. Rept.",
    volume = "1182",
    pages = "1--87",
    year = "2026"
}

@article{Bagchi:2010zz,
	Archiveprefix = {arXiv},
	Author = {Bagchi, Arjun},
	Doi = {10.1103/PhysRevLett.105.171601},
	Eprint = {1006.3354},
	Journal = {Phys. Rev. Lett.},
	Pages = {171601},
	Primaryclass = {hep-th},
	Title = {{Correspondence between Asymptotically Flat Spacetimes and Nonrelativistic Conformal Field Theories}},
	Volume = {105},
	Year = {2010}}

@article{Kim:2023qbl,
    author = "Kim, Seolhwa and Kraus, Per and Monten, Ruben and Myers, Richard M.",
    title = "{S-matrix path integral approach to symmetries and soft theorems}",
    eprint = "2307.12368",
    archivePrefix = "arXiv",
    primaryClass = "hep-th",
    doi = "10.1007/JHEP10(2023)036",
    journal = "JHEP",
    volume = "10",
    pages = "036",
    year = "2023"
}

@article{Kraus:2024gso,
    author = "Kraus, Per and Myers, Richard M.",
    title = "{Carrollian partition functions and the flat limit of AdS}",
    eprint = "2407.13668",
    archivePrefix = "arXiv",
    primaryClass = "hep-th",
    doi = "10.1007/JHEP01(2025)183",
    journal = "JHEP",
    volume = "01",
    pages = "183",
    year = "2025"
}

@article{Kraus:2025wgi,
    author = "Kraus, Per and Myers, Richard M.",
    title = "{Carrollian Partition Function for Bulk Yang-Mills Theory}",
    eprint = "2503.00916",
    archivePrefix = "arXiv",
    primaryClass = "hep-th",
    month = "3",
    year = "2025"
}

@article{Isen:2026xoc,
    author = "Isen, Jack and Kraus, Per and Monten, Ruben and Myers, Richard M.",
    title = "{The gravitational S-matrix from the path integral: asymptotic symmetries and soft theorems}",
    eprint = "2603.17045",
    archivePrefix = "arXiv",
    primaryClass = "hep-th",
    reportNumber = "CERN-TH-2026-044",
    month = "3",
    year = "2026"
}

@article{Bagchi:2026emg,
    author = "Bagchi, Arjun and Lipstein, Arthur and Mondal, Saikat and Zhang, Alex Jiayi",
    title = "{Carrollian ABJM: Fermions and Supersymmetry}",
    eprint = "2604.22582",
    archivePrefix = "arXiv",
    primaryClass = "hep-th",
    month = "4",
    year = "2026"
}

@article{Li:2021snj,
    author = "Li, Yue-Zhou",
    title = "{Notes on flat-space limit of AdS/CFT}",
    eprint = "2106.04606",
    archivePrefix = "arXiv",
    primaryClass = "hep-th",
    doi = "10.1007/JHEP09(2021)027",
    journal = "JHEP",
    volume = "09",
    pages = "027",
    year = "2021"
}

@article{Alday:2018pdi,
    author = "Alday, Luis F. and Bissi, Agnese and Perlmutter, Eric",
    title = "{Genus-One String Amplitudes from Conformal Field Theory}",
    eprint = "1809.10670",
    archivePrefix = "arXiv",
    primaryClass = "hep-th",
    doi = "10.1007/JHEP06(2019)010",
    journal = "JHEP",
    volume = "06",
    pages = "010",
    year = "2019"
}

@article{Alday:2024srr,
    author = "Alday, Luis F. and Zhou, Xinan",
    title = "{Flat-space limit of defect correlators and stringy AdS form factors}",
    eprint = "2411.04378",
    archivePrefix = "arXiv",
    primaryClass = "hep-th",
    doi = "10.1007/JHEP03(2025)182",
    journal = "JHEP",
    volume = "03",
    pages = "182",
    year = "2025"
}

@article{Chen:2025cod,
    author = "Chen, Qi and Tao, Yi-Xiao and Zhou, Xinan",
    title = "{Notes on flat-space limit of holographic defect correlators in position space}",
    eprint = "2507.14421",
    archivePrefix = "arXiv",
    primaryClass = "hep-th",
    doi = "10.1007/JHEP11(2025)123",
    journal = "JHEP",
    volume = "11",
    pages = "123",
    year = "2025"
}

@article{Goncalves:2014ffa,
    author = "Gon{\c{c}}alves, Vasco",
    title = "{Four point function of $\mathcal{N}=4$ stress-tensor multiplet at strong coupling}",
    eprint = "1411.1675",
    archivePrefix = "arXiv",
    primaryClass = "hep-th",
    doi = "10.1007/JHEP04(2015)150",
    journal = "JHEP",
    volume = "04",
    pages = "150",
    year = "2015"
}

@article{Alday:2014tsa,
    author = "Alday, Luis F. and Bissi, Agnese and Lukowski, Tomasz",
    title = "{Lessons from crossing symmetry at large N}",
    eprint = "1410.4717",
    archivePrefix = "arXiv",
    primaryClass = "hep-th",
    doi = "10.1007/JHEP06(2015)074",
    journal = "JHEP",
    volume = "06",
    pages = "074",
    year = "2015"
}

@article{Bagchi:2012cy,
	Archiveprefix = {arXiv},
	Author = {Bagchi, Arjun and Fareghbal, Reza},
	Doi = {10.1007/JHEP10(2012)092},
	Eprint = {1203.5795},
	Journal = {JHEP},
	Pages = {092},
	Primaryclass = {hep-th},
	Reportnumber = {EMPG-12-04, NI-12014},
	Slaccitation = {%%CITATION = ARXIV:1203.5795;%%},
	Title = {{BMS/GCA Redux: Towards Flatspace Holography from Non-Relativistic Symmetries}},
	Volume = {10},
	Year = {2012}}

@article{Barnich:2012aw,
	Archiveprefix = {arXiv},
	Author = {Barnich, Glenn and Gomberoff, Andres and Gonzalez, Hernan A.},
	Doi = {10.1103/PhysRevD.86.024020},
	Eprint = {1204.3288},
	Journal = {Phys. Rev. D},
	Pages = {024020},
	Primaryclass = {gr-qc},
	Title = {{The Flat limit of three dimensional asymptotically anti-de Sitter spacetimes}},
	Volume = {86},
	Year = {2012}}

@article{Bagchi:2012yk,
	Archiveprefix = {arXiv},
	Author = {Bagchi, Arjun and Detournay, Stephane and Grumiller, Daniel},
	Doi = {10.1103/PhysRevLett.109.151301},
	Eprint = {1208.1658},
	Journal = {Phys. Rev. Lett.},
	Pages = {151301},
	Primaryclass = {hep-th},
	Slaccitation = {%%CITATION = ARXIV:1208.1658;%%},
	Title = {{Flat-Space Chiral Gravity}},
	Volume = {109},
	Year = {2012}}

@article{Bagchi:2014iea,
      author         = "Bagchi, Arjun and Basu, Rudranil and Grumiller, Daniel
                        and Riegler, Max",
      title          = "{Entanglement entropy in Galilean conformal field
                        theories and flat holography}",
      journal        = "Phys. Rev. Lett.",
      volume         = "114",
      year           = "2015",
      number         = "11",
      pages          = "111602",
      doi            = "10.1103/PhysRevLett.114.111602",
      eprint         = "1410.4089",
      archivePrefix  = "arXiv",
      primaryClass   = "hep-th",
      reportNumber   = "TUW-14-14",
      SLACcitation   = "%%CITATION = ARXIV:1410.4089;%%"
}

@article{Jiang:2017ecm,
	Archiveprefix = {arXiv},
	Author = {Jiang, Hongliang and Song, Wei and Wen, Qiang},
	Doi = {10.1007/JHEP07(2017)142},
	Eprint = {1706.07552},
	Journal = {JHEP},
	Pages = {142},
	Primaryclass = {hep-th},
	Title = {{Entanglement Entropy in Flat Holography}},
	Volume = {07},
	Year = {2017}}

@article{LevyLeblond,
	Author = {Leblond, Levy},
	Journal = {Annales Poincare Phys.Theor.},
	Number = {1},
	Title = {{Une nouvelle limite non-relativiste du group de Poincar{\'e}}},
	Volume = {3},
	Year = {1965}}

@article{SenGupta:1966qer,
    author = "Sen Gupta, N. D.",
    title = "{On an analogue of the Galilei group}",
    doi = "10.1007/BF02740871",
    journal = "Nuovo Cim. A",
    volume = "44",
    number = "2",
    pages = "512--517",
    year = "1966"
}

@article{Duval:2014uva,
	Archiveprefix = {arXiv},
	Author = {Duval, C. and Gibbons, G. W. and Horvathy, P. A.},
	Doi = {10.1088/0264-9381/31/9/092001},
	Eprint = {1402.5894},
	Journal = {Class. Quant. Grav.},
	Pages = {092001},
	Primaryclass = {gr-qc},
	Slaccitation = {%%CITATION = ARXIV:1402.5894;%%},
	Title = {{Conformal Carroll groups and BMS symmetry}},
	Volume = {31},
	Year = {2014}}

@article{Bagchi:2024efs,
    author = "Bagchi, Arjun and Lipstein, Arthur and Mandlik, Mangesh and Mehra, Aditya",
    title = "{3d Carrollian Chern-Simons theory \& 2d Yang-Mills}",
    eprint = "2407.13574",
    archivePrefix = "arXiv",
    primaryClass = "hep-th",
    doi = "10.1007/JHEP11(2024)006",
    journal = "JHEP",
    volume = "11",
    pages = "006",
    year = "2024"
}

@article{Bondi:1962px,
	Author = {Bondi, H. and van der Burg, M. G. J. and Metzner, A. W. K.},
	Doi = {10.1098/rspa.1962.0161},
	Journal = {Proc. Roy. Soc. Lond. A},
	Pages = {21--52},
	Title = {{Gravitational waves in general relativity. 7. Waves from axisymmetric isolated systems}},
	Volume = {269},
	Year = {1962}}

@article{Sachs:1962wk,
	Author = {Sachs, R. K.},
	Doi = {10.1098/rspa.1962.0206},
	Journal = {Proc. Roy. Soc. Lond.},
	Pages = {103-126},
	Slaccitation = {%%CITATION = PRSLA,A270,103;%%},
	Title = {{Gravitational waves in general relativity. 8. Waves in asymptotically flat space-times}},
	Volume = {A270},
	Year = {1962}}

@article{Weinberg:2010fx,
    author = "Weinberg, Steven",
    title = "{Six-dimensional Methods for Four-dimensional Conformal Field Theories}",
    eprint = "1006.3480",
    archivePrefix = "arXiv",
    primaryClass = "hep-th",
    reportNumber = "UTTG-04-10",
    doi = "10.1103/PhysRevD.82.045031",
    journal = "Phys. Rev. D",
    volume = "82",
    pages = "045031",
    year = "2010"
}

@article{Costa:2011mg,
    author = "Costa, Miguel S. and Penedones, Joao and Poland, David and Rychkov, Slava",
    title = "{Spinning Conformal Correlators}",
    eprint = "1107.3554",
    archivePrefix = "arXiv",
    primaryClass = "hep-th",
    reportNumber = "LPTENS-11-22, NSF-KITP-11-128",
    doi = "10.1007/JHEP11(2011)071",
    journal = "JHEP",
    volume = "11",
    pages = "071",
    year = "2011"
}

@inproceedings{Simmons-Duffin:2016gjk,
    author = "Simmons-Duffin, David",
    title = "{The Conformal Bootstrap}",
    booktitle = "{Theoretical Advanced Study Institute in Elementary Particle Physics}: {New Frontiers in Fields and Strings}",
    eprint = "1602.07982",
    archivePrefix = "arXiv",
    primaryClass = "hep-th",
    doi = "10.1142/9789813149441_0001",
    pages = "1--74",
    year = "2017"
}

@article{Brown:2011pj,
    author = "Brown, Francis and Kreimer, Dirk",
    title = "{Angles, Scales and Parametric Renormalization}",
    eprint = "1112.1180",
    archivePrefix = "arXiv",
    primaryClass = "hep-th",
    doi = "10.1007/s11005-013-0625-6",
    journal = "Lett. Math. Phys.",
    volume = "103",
    pages = "933--1007",
    year = "2013"
}

@article{deGioia:2022fcn,
    author = "de Gioia, Leonardo Pipolo and Raclariu, Ana-Maria",
    title = "{Eikonal approximation in celestial CFT}",
    eprint = "2206.10547",
    archivePrefix = "arXiv",
    primaryClass = "hep-th",
    doi = "10.1007/JHEP03(2023)030",
    journal = "JHEP",
    volume = "03",
    pages = "030",
    year = "2023"
}

@article{deGioia:2023cbd,
    author = "de Gioia, Leonardo Pipolo and Raclariu, Ana-Maria",
    title = "{Celestial sector in CFT: Conformally soft symmetries}",
    eprint = "2303.10037",
    archivePrefix = "arXiv",
    primaryClass = "hep-th",
    doi = "10.21468/SciPostPhys.17.1.002",
    journal = "SciPost Phys.",
    volume = "17",
    number = "1",
    pages = "002",
    year = "2024"
}

@article{Cotler:2025npu,
    author = "Cotler, Jordan and Dhivakar, Prateksh and Jensen, Kristan",
    title = "{Carrollian Holographic Duals Are Nonlocal}",
    eprint = "2512.05072",
    archivePrefix = "arXiv",
    primaryClass = "hep-th",
    doi = "10.1103/f26s-4p77",
    journal = "Phys. Rev. Lett.",
    volume = "137",
    number = "9",
    pages = "091601",
    year = "2026"
}

@article{DiVecchia:2015jaq,
    author = "Di Vecchia, Paolo and Marotta, Raffaele and Mojaza, Matin and Nohle, Josh",
    title = "{New soft theorems for the gravity dilaton and the Nambu-Goldstone dilaton at subsubleading order}",
    eprint = "1512.03316",
    archivePrefix = "arXiv",
    primaryClass = "hep-th",
    reportNumber = "NORDITA-2015-134",
    doi = "10.1103/PhysRevD.93.085015",
    journal = "Phys. Rev. D",
    volume = "93",
    number = "8",
    pages = "085015",
    year = "2016"
}

@article{Borovik:2026zgl,
    author = "Borovik, Viktoriia and de Korte, Claire and Meurrens, Nathan and Pavlov, Dmitrii",
    title = "{Constraining Conformal Correlators}",
    eprint = "2605.31491",
    archivePrefix = "arXiv",
    primaryClass = "math.CO",
    month = "5",
    year = "2026",
}

@article{Donnay:2022sdg,
    author = "Donnay, Laura and Pasterski, Sabrina and Puhm, Andrea",
    title = "{Goldilocks modes and the three scattering bases}",
    eprint = "2202.11127",
    archivePrefix = "arXiv",
    primaryClass = "hep-th",
    reportNumber = "CPHT-RR004.012022",
    doi = "10.1007/JHEP06(2022)124",
    journal = "JHEP",
    volume = "06",
    pages = "124",
    year = "2022"
}

@article{Pasterski:2021dqe,
    author = "Pasterski, Sabrina and Puhm, Andrea and Trevisani, Emilio",
    title = "{Revisiting the conformally soft sector with celestial diamonds}",
    eprint = "2105.09792",
    archivePrefix = "arXiv",
    primaryClass = "hep-th",
    reportNumber = "CPHT-RR037.052021",
    doi = "10.1007/JHEP11(2021)143",
    journal = "JHEP",
    volume = "11",
    pages = "143",
    year = "2021"
}

@article{Puhm:2019zbl,
    author = "Puhm, Andrea",
    title = "{Conformally Soft Theorem in Gravity}",
    eprint = "1905.09799",
    archivePrefix = "arXiv",
    primaryClass = "hep-th",
    reportNumber = "CPHT-RR021.052019",
    doi = "10.1007/JHEP09(2020)130",
    journal = "JHEP",
    volume = "09",
    pages = "130",
    year = "2020"
}

@article{McLoughlin:2024ldp,
    author = "McLoughlin, Tristan and Moynihan, Nathan and Puhm, Andrea",
    title = "{Celestial amplitudes on electromagnetic backgrounds: T-duality from S-duality}",
    eprint = "2408.13234",
    archivePrefix = "arXiv",
    primaryClass = "hep-th",
    month = "8",
    year = "2024"
}

@unpublished{PV_Mukherjee:inprep,
  author = "{Athira P V} and {Ritwick Kumar Ghosh}",
  title  = "{Massless Fields at Spatial Infinity}",
  note   = "{In preparation}",
  year   = "2026"
}

@article{Haco:2017ekf,
    author = "Haco, Sasha J. and Hawking, Stephen W. and Perry, Malcolm J. and Bourjaily, Jacob L.",
    title = "{The Conformal BMS Group}",
    eprint = "1701.08110",
    archivePrefix = "arXiv",
    primaryClass = "hep-th",
    doi = "10.1007/JHEP11(2017)012",
    journal = "JHEP",
    volume = "11",
    pages = "012",
    year = "2017"
}

@article{Green:2020eyj,
    author = "Green, Michael B. and Wen, Congkao",
    title = "{Maximal U(1)$_{Y}$-violating n-point correlators in $ \mathcal{N} $ = 4 super-Yang-Mills theory}",
    eprint = "2009.01211",
    archivePrefix = "arXiv",
    primaryClass = "hep-th",
    reportNumber = "QMUL-PH-20-23",
    doi = "10.1007/JHEP02(2021)042",
    journal = "JHEP",
    volume = "02",
    pages = "042",
    year = "2021"
}

@article{Cotler:2025dau,
    author = "Cotler, Jordan and Dhivakar, Prateksh and Jensen, Kristan",
    title = "{A finite Carrollian critical point}",
    eprint = "2504.12289",
    archivePrefix = "arXiv",
    primaryClass = "hep-th",
    doi = "10.1007/JHEP08(2025)172",
    journal = "JHEP",
    volume = "08",
    pages = "172",
    year = "2025"
}

@article{Sen:2025oeq,
    author = "Sen, Ashoke",
    title = "{Decorating asymptotically flat space-time with the moduli space of string theory}",
    eprint = "2506.13876",
    archivePrefix = "arXiv",
    primaryClass = "hep-th",
    doi = "10.1007/JHEP12(2025)145",
    journal = "JHEP",
    volume = "12",
    pages = "145",
    year = "2025"
}

@article{Sen:2025ljz,
    author = "Sen, Ashoke",
    title = "{How to Create a Flat Ten or Eleven Dimensional Space-time in the Interior of an Asymptotically Flat Four Dimensional String Theory}",
    eprint = "2503.00601",
    archivePrefix = "arXiv",
    primaryClass = "hep-th",
    month = "3",
    year = "2025"
}

@article{Sen:2025bmj,
    author = "Sen, Ashoke",
    title = "{Are Moduli Vacuum Expectation Values or Parameters?}",
    eprint = "2502.07883",
    archivePrefix = "arXiv",
    primaryClass = "hep-th",
    month = "2",
    year = "2025"
}

@article{Pasterski:2016qvg,
    author = "Pasterski, Sabrina and Shao, Shu-Heng and Strominger, Andrew",
    title = "{Flat Space Amplitudes and Conformal Symmetry of the Celestial Sphere}",
    eprint = "1701.00049",
    archivePrefix = "arXiv",
    primaryClass = "hep-th",
    doi = "10.1103/PhysRevD.96.065026",
    journal = "Phys. Rev. D",
    volume = "96",
    number = "6",
    pages = "065026",
    year = "2017"
}
\bibliographystyle{JHEP}

\end{document}